\documentclass[11pt]{article}
\usepackage{latexsym} 
\usepackage{amsfonts}
\usepackage{amsmath}
\usepackage{amsthm}
\usepackage{amssymb}
\renewcommand{\theequation}{\thesection.\arabic{equation}}
\newcounter{subequation}[equation]
\makeatletter

\newcommand{\comment}[1]{}

\expandafter\let\expandafter\reset@font\csname reset@font\endcsname

\def\subeqnarray{\arraycolsep1pt
    \def\@eqnnum\stepcounter##1{\stepcounter{subequation}%
        {\reset@font\rm(\theequation\alph{subequation})}}
    \jot5mm     \eqnarray}

\makeatother

\newtheorem{proposition}{Proposition}

\def\be{\begin{equation}}
\def\ee{\end{equation}}

\def\lb{\label}

\def\bea{\begin{eqnarray}}
\def\eea{\end{eqnarray}}
\def\ba{\begin{array}}
    \def\ea{\end{array}}
\def\dd{\partial}

\def\half{\frac{1}{2}}

\def\one#1{#1^{\raise5pt\hbox{$\scriptstyle\!\!\!\!1$}}\,{}}
\def\two#1{#1^{\raise5pt\hbox{$\scriptstyle\!\!\!\!2$}}\,{}}

\def\tilde{\widetilde}
\def\II{\hbox{{1}\kern-.25em\hbox{l}}}
\def\a{\alpha}
\def\b{\beta}
\def\c{\gamma}
\def\d{\delta}
\def\e{\varepsilon}

\makeatletter
\def\binrel@#1{\begingroup
    \setboxz@h{\thinmuskip0mu
        \medmuskip\m@ne mu\thickmuskip\@ne mu
        \setbox\tw@\hbox{$#1\m@th$}\kern-\wd\tw@
        ${}#1{}\m@th$}%
    \edef\@tempa{\endgroup\let\noexpand\binrel@@
        \ifdim\wdz@<\z@ \mathbin
        \else\ifdim\wdz@>\z@ \mathrel
        \else \relax\fi\fi}%
    \@tempa
}
\let\binrel@@\relax
\def\overset#1#2{\binrel@{#2}%
    \binrel@@{\mathop{\kern\z@#2}\limits^{#1}}}
\def\underset#1#2{\binrel@{#2}%
    \binrel@@{\mathop{\kern\z@#2}\limits_{#1}}}
\makeatother
\newfont{\bbd}{msbm10 scaled\magstep1}

\begin{document}

\begin{center}

    {\LARGE {Spinorial $R$ operator and Algebraic Bethe Ansatz      }}

{\large \sf
D. Karakhanyan$^{a}$\footnote{\sc e-mail: karakhan@yerphi.am  },
R. Kirschner$^b$\footnote{\sc e-mail:Roland.Kirschner@itp.uni-leipzig.de}} \\

\vspace{0.5cm}

\begin{itemize}
\item[$^a$]
{\it Yerevan Physics Institute,
2 Alikhanyan br., 0036 Yerevan, Armenia }
\item[$^b$]
{\it Institut f\"ur Theoretische
Physik, Universit\"at Leipzig, \\
PF 100 920, D-04009 Leipzig, Germany}
\end{itemize}
\end{center}

\vspace{.5cm}
\begin{abstract}
\noindent
We propose a new approach to the spinor-spinor R-matrix with
orthogonal and symplectic symmetry. Based on this approach and 
 the fusion method we relate the
spinor-vector and vector-vector monodromy matrices for quantum spin  chains. 
We consider the explicit spinor $R$ matrices of low rank orthogonal algebras
and the corresponding $RTT$ algebras. Coincidences with fundamental $R$
matrices allow to relate the Algebraic Bethe Ansatz for spinor and vector
monodromy matrices. 
\end{abstract}

\vspace{2cm}

   \renewcommand{\refname}{References.}
   \renewcommand{\thefootnote}{\arabic{footnote}}
   \setcounter{footnote}{0}

\newpage

\section{Introduction}

The Inverse Scattering Method for quantum integrable systems
\cite{FST,TTF,KuSk1,Fad}
applies basically to symmetries of all Lie algebras types. The case of the $A$
series has been studied in much detail, because of its applications to well
known models and because of its simplicity compared to other cases. In last
years the Yangian algebras of the $B, C, D$ types and the corresponding
Algebraic Bethe Ansatz (ABA) attracted increasing interest, e.g.
\cite{AMR,CDI,JLM,GRW,BN,lprs,jly}. 

Yangians of the orthogonal and symplectic types allow for the linear
evalutation only by additional constraints on the universal enveloping of
the corresponding Lie algebra. The spinor representations, where the Yangian
generators are constructed from the underlying algebra $\mathcal{C}$ (see
eq.(\ref{cab}) below), obey these constraints, i.e. the Yang Baxter $RLL$ relation
involving the fundamental $R$ matrix with orthogonal or symplectic
symmetry and the spinor $L$ operators holds \cite{R85}. 

Reshetikhin \cite{r} proposed to approach the ABA in the orthogonal case by replacing
in the auxiliary space the fundamental (vector) representation by the spinor
representation and to apply the fusion procedure. The ABA for the $so(3)$
case has been shown to be related to the well known $s\ell(2)$ ABA. 
Our investigation is oriented along this concept.

We consider the Yang-Baxter $RLL$ relation in the tensor product of the
fundamental and two copies of the spinor representations. It is
used as the defining relation for the involved $R$ operator intertwining the
spinor representations. The spinorial $R$ operator has been obtained in the
orthogonal case in \cite{sw} and considered in \cite{R85, CDI}. The uniform
formulation of the orthogonal and symplectic cases and of the superalgebra 
case have been given in \cite{IKK}. The result of the spinorial $R$
operator has been given in the expansion in invariants on the tensor
product of two copies of the algebra $\mathcal{C}$ built as contraction of 
anti-symmetrized (orthogonal case) or symmetrized (symplectic case) products
of its generators.  

In this paper we propose an alternative approach. We use instead the
invariant built from the product of one generator out of each of the
spinor algebras and its powers.  We obtain the
spinorial $R$ operator in terms of the Euler Beta function.
We consider  low rank orthogonal cases. Here the spinor representations
are finite dimensional. This results in characteristic polynomials obeyed
by the invariant from which the explicit form of projection operators can be
read off. The general result of the spinorial $R$ operator is reduced to
a finite sum, which is derived from the spectral decomposition of  the
invariant. 

We observe coincidences of the spinorial  and  known fundamental (vector) $R$
matrices. 
The structure of the spinorial $R$ in even dimensions ($so(2m)$, $D$
series) appears much simpler than the odd dimensional case ($so(2m+1)$,
$B$ series). 

In the even-dimensional cases  the separation of the tensor product of
spinor representation spaces, where $R$ acts, into chiral parts 
results in the corresponding separation of $R$. Both parts of $R$ obey the
Yang-Baxter relations separately. 
The explicit matrix forms of $R$ have a lot of zero entries. This also
appears in the corresponding monodromy matrices. 

The observed coincidences of the spinorial $R$ matrices imply 
coincidences of the corresponding $RTT$ (spinorial Yangian) algebras and
fundamental $RTT$ (ordinary Yangian) algebras and as a consequence
relations of the corresponding ABA. 

In particular we observe the coincidences of the spinorial $so(4)$ with fundamental
$s\ell(2) \oplus s\ell(2) $ and the spinorial $so(6)$ with fundamental $s\ell(4)
\oplus s\ell(4) $ $R$ matrices. 
Besides of the well known relations between spinorial $so(3)$,
fundamental $s\ell(2)$ and fundamental $sp(2)$ Yangian algebras we 
observe the equivalence of the spinorial $so(5)$ and the fundamental 
$sp(4)$ Yangian algebras.

The paper is organized as follows: 

In sect. 2 we recall the Yang-Baxter relations with orthogonal and symplectic
symmetry, in particular  the ones involving the fundamental and the spinor
representations. In sect. 3 we describe our approach to  
the spinor-spinor $R$-matrix and compare with the Shankar-Witten approach. 
In sect. 4 we consider the relation between spinorial and vector monodromy
matrices using the fusion procedure. In sect. 5 we analyze the  spinorial
$R$ matrices for $so(4)$ and $so(6)$ and the generated $RTT$ algebras.
The spinorial $R$ matrices for $so(3)$ and $so(5)$ are analyzed in sect. 6.

\section{The spinorial $R$-matrix}
\setcounter{equation}{0}

The  Yang-Baxter relation for the fundamental $R$ matrix 
reads
\be\lb{rrr}
R^{a_1a_2}_{b_1b_2}(u-v)R^{b_1a_3}_{c_1b_3}(u)R^{b_2b_3}_{c_2c_3}(v)=
R^{a_2a_3}_{b_2b_3}(v)R^{a_1b_3}_{b_1c_3}(u)R^{b_1b_2}_{c_1c_2}(u).
\ee
 $R$ looks 
similar for the cases of orthogonal and symplectic symmetry\cite{z}, \cite{ka},
 \be\label{rzW}
R_{b_1b_2}^{a_1a_2}(u)=u(u + \frac{n}{2} -\epsilon)I^{a_1a_2}_{b_1b_2}
+(u + \frac{n}{2} -\epsilon)P^{a_1a_2}_{b_1b_2}-\epsilon \, u \, K^{a_1a_2}_{b_1b_2} \; ,
 \ee
 where
 \be\lb{KP}
I^{a_1a_2}_{b_1b_2} = \delta^{a_1}_{b_1} \delta^{a_2}_{b_2}   \, , \qquad
P^{a_1a_2}_{b_1b_2} = \delta^{a_1}_{b_2} \delta^{a_2}_{b_1}   \, , \qquad
K^{a_1a_2}_{b_1b_2} = \varepsilon^{a_1a_2} \, \varepsilon_{b_1b_2} .
\ee
 The choices $\epsilon = +1$ and $\epsilon = -1$ correspond to the 
$so(n)$ and $sp(2m)$ cases respectively. The index range is
$a_i, b_i = 1,...,n$ or $1,...,2m$.
$\e^{ab}$ denotes the metric tensor which is symmetric in the orthogonal
and anti-symmetric in the symplectic case.

The Yang- Baxter $RLL$ relation with the above $R$
\be\lb{rll1}
R^{a_1a_2}_{b_1b_2}(u)L^{b_1}_{c_1}(u+v)L^{b_2}_{c_2}(v)=
L^{a_2}_{b_2}(v)L^{a_1}_{b_1}(u+v)R^{b_1b_2}_{c_1c_2}(u),
\ee
is fulfilled by the linear form of the  $L$-operator
\be\lb{losc}
L^a{}_b(u)=u\d^a_b+G^a{}_b,
 \ee
in the spinor representation case, where the matrix elements  $G^a_b$ 
are built as
\be \label{Gcc}
G^a_b = \half (c^a c_b- \epsilon c_b c^a) 
\ee
from the underlying algebra $\mathcal{C}$ generated by $c_a$
obeying the commutation relations
 \be\lb{cab}
c^ac^b+\epsilon c^bc^a=\e^{ab},\qquad{\rm{or}}\qquad (c^a)^\a_
\c(c^b)^\c_\b+\epsilon(c^b)^a_\c(c^a)^\c_\b=\e^{ab}\d^\a_\b.
 \ee
 
The linear ansatz for $L$-operator (\ref{losc})
 implies the $so(n)$ or $sp(2m)$ Lie algebra relations,
 \be\lb{algg}
[G^{a_1}{}_{b_1},G^{a_2}{}_{b_2}]=\d^{a_1}_{b_2}G^{a_2}{}_{b_1}-\d^
{a_2}{}_{b_1}G^{a_1}{}_{b_2}+\epsilon\e^{a_1a_2}\e_
{b_1c_2}G^{c_2}{}_{b_2}-\epsilon G^{a_2}{}_{c_2}\e^{a_1c_2}\e_{b_1b_2},
 \ee
and the symmetry condition,
 \be\lb{sc}
\e^{a_1a_2}\e_{c_1c_2}(\d^{c_1}_{b_1}G^{c_2}{}_{b_2}+\d^{c_2}_{b_2}
G^{c_1}{}_{b_1})=(\d^{a_1}_{c_1}G^{a_2}{}_{c_2}+\d^{a_2}_{c_2}G^{a_1}
{}_{c_1})\e^{c_1c_2}\e_{b_1b_2},
 \ee
as well as the additional constraint
 \be\lb{ar}
\e^{a_1a_2}\e_{c_1c_2}(G^{c_1}{}_{b_1}-\b\d^{c_1}_{b_1})
G^{c_2}{}_{b_2}=G^{a_2}{}_{c_2}(G^{a_1}
{}_{c_1}-\b\d^{a_1}_{c_1})\e^{c_1c_2}\e_{b_1b_2},
 \ee
which specifies the Yangian linear evaluation.

The Yang-Baxter relation is formulated in the tensor product of
 three  representation spaces. We are concerned with the
fundamental (vector) space $\mathcal{V}$ ($2m$- or $2m+1$- dimensional)
and the spinor space $\mathcal{S}$ (of dimension $2^m$ in the
orthogonal and infinite dimensional, oscillator Fock space, in the symplectic
case).

 In order to specify the representation involved  we shall show all indices explicitly,
$a,b$ for $\mathcal{V}$ and $\alpha, \beta$ for $\mathcal{S}$.
Then  (\ref{rll1}) appears as 
\be\lb{rll2}
R^{a_1a_2}_{b_1b_2}(u)L^{b_1\a}_{c_1\c}(u+v)L^{b_2\c}_{c_2\b}(v)=
L^{a_2\a}_{b_2\c}(v)L^{a_1\c}_{b_1\b}(u+v)R^{b_1b_2}_{c_1c_2}(u).
\ee
 Along with this Yang-Baxter relation we consider 
\be\lb{rll3}
{\mathfrak{R}}^{\a_1\a_2}_{\b_1\b_2}(u)L^{a\b_1}_{c\c_1}(u+v)
L^{c\b_2}_{b\c_2}(v)=L^{a\a_2}_{c\b_2}(v)L^{c\a_1}_{b\b_1}(u+v)
{\mathfrak{R}}^{\b_1\b_2}_{\c_1\c_2}(u).
\ee
Here ${\mathfrak{R}}(u)$ stands for the spinorial $R$-operator.

We shall use also the modifications of the above $RLL$ relations usually
referred to as the check form. The $R$ matrix is multiplied by the
permutation operator interchanging the auxiliary space factors, 
$\check R_{12} = \mathcal{P}_{12} R_{12}$. For example, the check form
corresponding to (\ref{rll3}) is 
\be \label{rll3c}
\check {\mathfrak{R}}^{\a_1\a_2}_{\b_1\b_2}(u)L^{a\b_1}_{c\c_1}(u+v)
L^{c\b_2}_{b\c_2}(v)=L^{a\a_1}_{c\b_1}(v)L^{c\a_2}_{b\b_2}(u+v)
\check {\mathfrak{R}}^{\b_1\b_2}_{\c_1\c_2}(u).
\ee  

In \cite{IKK}, following the approach of \cite{sw}, the spinorial 
$R$-operator was obtained in the form
\be \label{Rspin}
\check{\mathfrak{R}}_{12}(u) = \sum_k \frac{r_k(u)}{k!} 
\sum_{\vec{a},\vec{b}}\e_{a_1b_1}\ldots\e_{a_kb_k} c_1
^{[a_1}\ldots c_1^{a_k)} \cdot c_2^{[b_1}\ldots c_2^{b_k)}. 
\ee
Here $c_1^{[a_1} \ldots c_1^{a_k)}$ symbolizes the anti-symmetrization in the
orthogonal and the symmetrization in the symplectic case.
The coefficients are derived from an iterative relation as
$$ 
r_{2m} = \frac{2^{2m} \Gamma(m+\e \frac{u}{2})}{\Gamma(m+1-\e
\frac{u+n}{2}) } A_0(u),
$$
$$ 
r_{2m+1} = \frac{2^{2m} \Gamma(m+\half + \e \frac{u}{2})}
{\Gamma(m+\half-\e \frac{u+n}{2}) }A_1(u),
$$ 
where $A_0$ and $A_1$ are arbitrary functions of the spectral parameter.

Both the even (sum restricted to even $k$) and the odd parts obey 
(\ref{rll3c}).

Note that in the orthogonal case $c^a$, the generators of $\mathcal{C}$,
 are fermionic and can be expressed in terms of the
 Dirac gamma matrices, $c^a=\frac1{\sqrt{2}}\c^a$.

\section{The alternative approach to  ${\check{\mathfrak{R}}}$ }
\setcounter{equation}{0}

The spinorial $\mathfrak{R}$ operator  can be regarded as an element of
$\mathcal{C}_1 \otimes \mathcal{C}_2 $ (\ref{cab}) or as an operator 
acting in the product of 
 two copies of the spinor space, $\mathcal{S}_1\otimes\mathcal{S}_2$.  
$c^a_1$ and $c^a_2$ (\ref{cab}) are the corresponding basic elements.

The equation (\ref{rll3c})  takes the form
 \be\lb{rllc3}
\check{\mathfrak{R}}_{12}(u)\Big((u+v+\frac\epsilon2)\d^a_e-c^a_1c_{1e}\Big)
\Big((v+\frac\epsilon2)\d^e_b-c^e_2c_{2b}\Big)=
 \ee
$$
=\Big((v+\frac\epsilon2)\d^a_e
-c^a_1c_{1e}\Big)\Big((u+v+\frac\epsilon2)\d^e_b-c^e_2c_{2b}\Big)
\check{\mathfrak{R}}_{12}(u).
$$
We consider it as the defining relation for the wanted spinorial $R$ matrix
and expand in powers of   $v$. 
The terms proportional
to $v^2$ are canceled, linear terms lead to the symmetry condition
 \be\lb{sym1}
\check{\mathfrak{R}}_{12}(u)(c^a_1c_{1b}+c^a_2c_{2b})=
(c^a_1c_{1b}+c^a_2c_{2b})\check{\mathfrak{R}}_{12}(u),
 \ee
from which one deduces that $\check{\mathfrak{R}}_{12}(u)$ depends
on  $c_1$ and $c_2$ through the invariant combination 
 \be\lb{z}
z=-iz_{12}=-ic^a_1\e_{ab}c^b_2=-i\epsilon c_{1e}c^e_2. 
 \ee
The last relation from  $v^0$  has the form
 \be\lb{def1}
\check{\mathfrak{R}}_{12}(u)(-uc^a_2c_{2b}+c^a_1z_{12}c_{2b})=
(-uc^a_1c_{1b}+c^a_1z_{12}c_{2b})\check{\mathfrak{R}}_{12}(u).
 \ee
Introducing 
$
c_{\pm}^a=c_1\pm ic_2$, 
we have the important relation for the following calculation
\be \label{zcpm} 
zc^a_\pm=c^a_\pm(z\pm1). \ee

Consider first the symmetry condition (\ref{sym1}). 
  Multiplying it by $c^d_\pm\e_{da}$ from the left and 
by $c^b_\pm$ from the right one obtains
$$
\check{\mathfrak{R}}_{12}(u|z\mp1)A=A\check{\mathfrak{R}}_{12}(u|z\pm1),
$$ 
where
$$
A=c^d_\pm\e_{da}(c^a_1c^b_1+c^a_2c^b_2)\e_{eb}c^e_\pm=
(\epsilon\frac d2\pm z)(\frac d2\mp\epsilon z)+(\pm i\epsilon\frac d2+iz)
(\pm i\frac d2-i\epsilon z)=0,
$$ 
here $d\equiv\e_{ba}\e^{ab}=\d^a_a$.

For the analogous projection with the opposite choice of signs one has 
$$
\check{\mathfrak{R}}_{12}(u|z\mp1)B=B\check{\mathfrak{R}}_{12}(u|z\mp1),
$$ 
where
$$
B=c^d_\pm\e_{da}(c^a_1c^b_1+c^a_2c^b_2)\e_{eb}c^e_\mp=
(\epsilon\frac d2\pm z)(\frac d2\pm\epsilon z)+(\pm i\epsilon\frac d2+iz)
(\mp i\frac d2-i\epsilon z)=2\epsilon(\frac d2\pm\epsilon z)^2.
$$

The analogous consideration of the
defining relation (\ref{def1})  leads to the following equation for the opposite
signs,
$$
\check{\mathfrak{R}}_{12}(u|z\mp1)C_L=
C_R\check{\mathfrak{R}}_{12}(u|z\mp1),
$$ 
where
$$
C_L=c^d_\pm\e_{da}(-uc^a_2c^b_2+c^a_1i\epsilon zc^b_2)\e_{eb}c^e_\mp=
$$
$$
=-u(\pm i\epsilon\frac d2+iz)(\mp i\frac d2-i\epsilon z)+(\epsilon\frac d2
\pm z)i\epsilon z(\mp i\frac d2-i\epsilon z)=(-u\epsilon\pm z)(\frac d2\pm\epsilon z)^2,
$$ 
$$
C_R=c^d_\pm\e_{da}(-uc^a_1c^b_1+c^a_1z_{12}c^b_2)\e_{eb}c^e_\mp=
$$
$$
=-u(\epsilon\frac d2\pm z)(\frac d2\pm\epsilon z)+(\epsilon\frac d2\pm z)
i\epsilon z(\mp i\frac d2-i\epsilon z)=(-u\epsilon\pm z)(\frac d2\pm\epsilon z)^2.
$$ 
The projection of  (\ref{def1}) with the same signs leads to
$$
\check{\mathfrak{R}}_{12}(u|z\mp1)D_L=D_R\check{\mathfrak{R}}_{12}(u|z\pm1),
$$ 
where
$$
D_L=c^d_\pm\e_{da}(-uc^a_2c^b_2+c^a_1i\epsilon zc^b_2)\e_{eb}c^e_\pm=
$$
$$
=-u(\pm i\epsilon\frac d2+iz)(\pm i\frac d2-i\epsilon z)+(\epsilon\frac d2
\pm z)i\epsilon z(\pm i\frac d2-i\epsilon z)=(u\epsilon\mp z)(\frac{d^2}4-z^2),
$$ 
$$
D_R=c^d_\pm\e_{da}(-uc^a_1c^b_1+c^a_1z_{12}c^b_2)\e_{eb}c^e_\pm=
$$
$$
=-u(\epsilon\frac d2\pm z)(\frac d2\mp\epsilon z)+(\epsilon\frac d2
\pm z)i\epsilon z(\pm i\frac d2-i\epsilon z)=(-u\epsilon\mp z)(\frac{d^2}4-z^2).
$$ 
 Canceling the common factor $\frac{d^2}4-z^2$ in both sides one obtains
 \be\lb{rc1}
\check{\mathfrak{R}}_{12}(u|z\mp1)(u\epsilon\mp z)=
(-u\epsilon\mp z)\check{\mathfrak{R}}_{12}(u|z\pm1),
 \ee 
from which one deduces the wanted spinorial $R$ operator
 \be\lb{rc2}
\check{\mathfrak{R}}_{12}(u|z)=r(u)\frac{\Gamma(\frac12(z+1-u\epsilon))}
{\Gamma(\frac12(z+1+u\epsilon))}.
 \ee

\subsection{Comparison with the Shankar-Witten form}

Comparing to the previous treatments \cite{sw,CDI,IKK}
we have a simpler line of arguments and a compact general form
of the spinorial $R$ operator.

The direct comparison of the two expressions for spinorial $R$-operator seems 
difficult due to the complicated connection between the invariants $I_k=\frac1{k!}
\e_{a_1b_1}\ldots \e_{a_kb_k} c_1^{[a_1}\ldots c_1^{a_k)} \cdot 
c_2^{[b_1}\ldots c_2^{b_k)}$ and $z$.
\be\lb{imz}
I_{k+1}=zI_k-\frac k4\big(d-\epsilon(k-1)\big)I_{k-1},\qquad\qquad
I_0(z,d)=1,\qquad I_1(z,d)=z.
\ee
Again the difference between orthogonal and symplectic cases consists in the
presence of the sign factor $\epsilon$. In the orthogonal case $so(d)$ ($\epsilon=+1$)  
the series in $I_k$ terminates at $k=d+1$, while for $sp(d)$  
($\epsilon= -1$)  the series does not terminate.

For a proof of the recurrence relation (\ref{imz}) we refer to
\cite{IKK}, where the generating function method is used for the
calculation of (anti-) symmetrized products of $c^a$. It 
follows the same line as for eq. (5.6) of that paper:
$$
zI_m=c_1^{[a_1}\ldots c_1^{a_m)}c_2^{[a_1}\ldots c_2^{a_m)}c_1^{a}
c_2^{a}=\dd_{\kappa_1}^{a_1\ldots a_m}\dd_{\kappa_2}^{a_1\ldots 
a_m}(\dd_{\kappa_1}^a+\frac12\kappa_1^a)(\dd_{\kappa_2}^a+
\frac12\kappa_2^a)e^{\kappa_1\cdot c_1+\kappa_2\cdot c_2}\Big|
_{\kappa_i=0}=
$$
$$
=I_{m+1}+\frac m4\big(d-(m-1)\epsilon\big)I_{m-1}.
$$

\subsection{Symplectic cases of low rank}

It is instructive to specify 
the general solution for spinorial $R$-matrix  for symplectic
algebras of low rank. For $sp(2)$ one can realize the  algebra $\mathcal{C}$
in terms of a pair of operators of multiplication and differentiation as
$$
c_1=\partial=c^{-1},\qquad c_{-1}=x=-c^1,\qquad z=-i(x_1\partial_2-x_2\partial_1).
$$
Thus we have
$$
\check{\mathfrak{R}}^{sp(2)}_{12}(u)=r(u)\frac{\Gamma\big(\frac{u+1}2
-\frac i2(x_1\dd_2-x_2\dd_1)\big)}
{\Gamma\big(\frac{1-u}2-\frac i2(x_1\dd_2-x_2\dd_1)\big)}.
$$
The $sp(4)$ case corresponds to two such pairs. 
$$
c^1=c^\dagger=x,\quad c^2=d^\dagger=y,\quad c^{-1}=\partial_x,\quad c^{-2}=
\partial_y
$$
$$
z=-i(x_1\partial_{x_2}-x_2\partial_{x_1}+y_1\partial_{y_2}-y_2
\partial_{y_1}).
$$
Thus we have
$$
\check{\mathfrak{R}}^{sp(4)}_{12}(u)=r(u)\frac{\Gamma\big
(\frac{u+1}2-\frac i2(x_1\dd_{x_2}-x_2\dd_{x_1}-\frac i2(y_1
\dd_{y_2}-y_2\dd_{y_1}))\big)}{\Gamma\big(\frac{1-u}2-\frac 
i2(x_1\dd_2-x_2\dd_1-\frac i2(y_1\dd_{y_2}-y_2\dd_{y_1}))\big)}.
$$


\subsection{The orthogonal case}

In the orthogonal case, due to the Clifford relation for  $c^a$  (Dirac gamma
matrices), the general expression for the spinor-spinor $R$-matrix in form of
the Euler
Beta-function can be transformed to a polynomial in $z$ as well as to a finite
expansion in the invariants
\be\lb{ik}
I_k=\c^{a_1\ldots a_k}\c^{a_1\ldots a_k},\qquad\qquad\qquad k=0,1,\ldots n.
\ee
In this case it is convenient to  modify slightly the definition of the invariant $z$:
\be\lb{zo}
z=\frac12\gamma_1^a\gamma_2^a.
\ee 
Besides of the absence of the imaginary unit in this definition, we suppose
here that the Dirac gamma matrices $\c^a_1$ and $\c^b_2$ related to 
different spaces {\it{commute}}, in contrast to the
previous convention where we have supposed their {\it{anticommutation}} in order
to have the unified description with the symplectic case.

Having modified the definition of $z$ we go through the steps of the above
derivation in order to check that the the result is not changed.

Consider  the RLL-relation 
$$
\check{\mathfrak{R}}_{12}(u|z)\Big((v+\frac12)\gamma_1^a\gamma_{1b}+(u+v+
\frac12)\gamma_2^a\gamma_{2b}-\gamma_1^az\gamma_{2b}\Big)=
$$
$$
=\Big((v+\frac12)\gamma_1^a\gamma_{1b}+(u+v+\frac12)\gamma_2
^a\gamma_{2b}-\gamma_1^az\gamma_{2b}\Big)\check{\mathfrak{R}}_{12}(u|z).
$$
We multiply by
$\gamma^a_\pm=\gamma^a_1\pm\gamma^2_2$ from the left
and by $\gamma^b_\pm$ or  $\gamma^b_\mp$ from the right and
use 
\be\lb{ccr}
z\gamma^a_\pm=\c^a_\pm(-z\pm1).
\ee
Note that here an additional minus sign appears due the change
in the commutativity convention. 
We obtain
$$
\check{\mathfrak{R}}_{12}(u|-z\pm1)(u+2v+1\mp z)(d\pm2z)^2
=(u+2v+1\mp z)(d\pm z)^2\check{\mathfrak{R}}_{12}(u|-z\pm1),
$$ 
and
$$
\check{\mathfrak{R}}_{12}(u|-z\pm1)(-u\pm\frac12z)(d^2-4z^2)
=(u\pm\frac12z)(d^2-4z^2)\check{\mathfrak{R}}{12}(u|-z\mp1),
$$
from which we deduce the  general solution for the $so(n)$ case:
\be\lb{sons}
\frac{\check{\mathfrak{R}}_{12}(u|z)}{{\mathfrak{R}}_0(u)}=\frac
{\Gamma(\frac{z+1-u}2)\Gamma(2u)}{\Gamma(\frac{z+1+u}2)}
=B\big(\frac{z+1-u}2,u\big).
\ee

\section{Spinor and vector monodromy matrices}
\setcounter{equation}{0}

We show how the monodromy matrices with spinor auxiliary space
($\mathcal{S}_0$) and vector auxiliary space ($\mathcal{V}_0$) 
are related by fusion. The gamma matrices are known to intertwine
the vector  and the product of spinor representations. 
Formally, the following analysis can be
extended to the symplectic case. However, that case is connected with the problem
of defining  infinite sums over the spinor indices. We restrict 
ourselves to the orthogonal case, which will be considered in the 
remaining part of the paper.

The general vector-vector monodromy matrix is defined by the 
fundamental $R$ matrix (\ref{rzW}):
\be\lb{tvv}
{\mathbb{T}}^{a_0,a_1\ldots a_N}_{\;\;\;\;b_0,b_1\ldots b_N}(u)=
R^{a_0a_1}_{a'_1b_1}(u)\ldots R^{a'_{N-1}a_N}_{b_0b_N}(u).
\ee
We are interested in the diagonalization of the trace of this matrix,
\be\lb{tr}
{\mathrm{t}}^{a_1\ldots a_N}_{\;\;\;\;b_1\ldots b_N}(u)={\mathbb{T}}^
{a_0,a_1\ldots a_N}_{\;\;\;\;a_0,b_1\ldots b_N}(u),
\ee
or ${\mathrm{t}}(u)=tr_0{\mathbb{T}}(u)$, because it is the generating function 
of the integrals of motion of a periodic quantum spin chain.

Using the $L$-operator (\ref{losc}) one can construct another
monodromy matrix
\be\lb{tv}
T^{\a,a_1\ldots a_N}_{\;\;\;\;\b,b_1\ldots b_N}(u)=\big(L^{a_1}{}
_{b_1}(u)\ldots L^{a_N}{}_{b_N}(u)\big)^\a{}_\b,
\ee
acting in the tensor product of the same quantum space $\mathcal{V}_1\otimes
\ldots\otimes \mathcal{V}_N$ and the spinor auxiliary space $\mathcal{S}_0$.

We consider also the related monodromy matrix
$$
\tilde T^{\c,c_1\ldots c_N}_{\;\;\;\;\d,b_1\ldots b_N}(-u-\frac12)
=\Big(L^{c_N}{}_{b_N}(-u-\frac12)L^{c_{N-1}}{}_{b_{N-1}}(-u-\frac12)
\ldots L^{c_1}{}_{b_1}(-u-\frac12)\Big)^\c_\d .
$$
Because of the inversion relation,
\be \label{invL}
L^{a\a}_{b\b}(u+\b+\frac12)L^{b\b}_{c\c}(-u-\frac12)=-u(u+\b+1)\d^a_c
\d^\a_\c,
\ee
it can be expressed  by the inverse of the previous monodromy matrix
$$
\tilde T^{\c,c_1\ldots c_N}_{\;\;\;\;\d,b_1\ldots b_N}(-u-\frac12) =
=\Big(-u(u+\b+1)\Big)^N\Big(T^{-1}(u+\b+\frac12)
\Big)^{\c,c_1\ldots c_N}_{\;\;\;\;\d,b_1\ldots b_N}.
$$ 
It is not hard to check that the fusion of two conjugated 
$L$-operators (\ref{losc}) gives the vector $R$-matrix (\ref{rzW}).
Indeed, 
$$
L^{a_1\a}_{c_1\b}(u+\b-\frac12)(\c_{b_0})^{\b}_{\c}L^{c_1\c}_{b_1\d}
(-\frac12-u)=-\Big(u(u+\b)\d^{a_1}_{b_1}\d^{a_0}_{b_0}+
(u+\b)\d^{a_1}_{b_0}\d^{a_0}_{b_1}-u\d^{a_1a_0}\d_{b_1b_0}
\Big)(\c_{a_0})^{\a}_{\d} =
$$ \be \label{LLR}
= -R^{a_0a_1}_{b_0b_1}(u)
(\c_{a_0})^{\a}_{\d}. 
\ee
We have used that in the orthogonal case (\ref{losc}) takes the form
\be\lb{llt}
L^{ab}(u)=u\d^{ab}-\frac12\c^{ab}.
\ee
Consider the product
$$
T^{\a,a_1\ldots a_N}_{\;\;\;\;\b,c_1\ldots c_N}(u+\b-\frac12)(\c_{b_0})
^{\b}_{\c}\tilde T^{\c,c_1\ldots c_N}_{\;\;\;\;\d,b_1\ldots b_N}(-u-\frac12)
=
$$
$$
 L^{\a,a_1}_{\a'_1,c_1}(u+\b-\frac12)L^
{\a'_1,a_2}_{\a'_2,c_2}(u+\b-\frac12)\ldots L^{\a'_{N-1},a_N}_{
\b,c_N}(u+\b-\frac12)\times.
$$
$$
\times (\c_{b_0})^{\b}_{\c}L^{\c,c_N}_{\c'_{N-1},b_N}(-\frac12-u)L^
{\c'_{N-1},c_{N-1}}_{\c'_{N-2},b_{N-1}}(-\frac12-u)\ldots L^{\c'_1,c_
1}_{\d,b_1}(-\frac12-u)
$$

We pick the last $L$ factor in $T$ and the first in $\tilde T$ and use 
(\ref{LLR}) to continue the calculation
\be\lb{ttt}
=L^{\a,a_1}_{\a'_1,c_1}(u+\b-\frac12)L^{\a'_1,a_2}_{\a'_2,c_2}(u+\b-
\frac12)\ldots L^{\a'_{N-2},a_{N-1}}_{\a'_{N-1},c_{N-1}}(u+\b-\frac12)
(\c^{a_0})^{\d}_{\a}\times
\ee
$$
\times\Big(L^{\a'_{N-1},a_N}_{\b,c_N}(u+\b-\frac12)(\c_{b_0})^{\b}_
{\c}L^{\c,c_N}_{\c'_{N-1},b_N}(-\frac12-u)\Big)\times
$$
$$
\times L^{\c'_{N-1},c_{N-1}}_{\c'_{N-2},b_{N-1}}(-\frac12-u)\ldots 
L^{\c'_2,c_2}_{\c'_1,b_2}(-\frac12-u)L^{\c'_1,c_1}_{\d,b_1}(-\frac12-u)=
$$
$$
=L^{\a,a_1}_{\a'_1,c_1}(u+\b-\frac12)L^{\a'_1,a_2}
_{\a'_2,c_2}(u+\b-\frac12)\ldots L^{\a'_{N-2},a_{N-1}}_{\a'_{N-1},c_{N
-1}}(u+\b-\frac12)\times
$$
$$
\times\Big(-R^{c_0a_N}_{b_0b_N}(u)(\c_{c_0})^{\a'_{N-1}}_{\c'_{N-1}}
\Big)\times
$$
$$
\times L^{\c'_{N-1},c_{N-1}}_{\c'_{N-2},b_{N-1}}(-\frac12-u)\ldots 
L^{\c'_2,c_2}_{\c'_1,b_2}(-\frac12-u)L^{\c'_1,c_1}_{\d,b_1}(-\frac12-u)=
$$
Proceeding with the next factors in $T$ and $\tilde T $  we obtain finally
$$
=(-1)^N{\mathbb{T}}^{a_0,a_1\ldots a_N}_{\;\;\;\;b_0,b_1
\ldots b_N}(u)(\c_{a_0})_{\d}^{\a}.
$$
In this way we obtain the fusion relation between spinorial and vector
monodromy matrices. The fusion of two spinorial monodromy matrices
by trace over the auxiliary spinor space $\mathcal{S}_0$  results in the
 vector monodromy matrix with $\mathcal{V}_0$ as auxiliary space.
\be\lb{t2t}
T^{\a,a_1\ldots a_N}_{\;\;\;\;\b,c_1\ldots c_N}(u+\b-\frac12)(\c_{b_0})
^{\b}_{\c}(T^{-1})^{\c,c_1\ldots c_N}_{\;\;\;\;\d,b_1\ldots b_N}(u+\b+\frac12)
={\mathbb{T}}^{a_0,a_1\ldots a_N}_{\;\;\;\;b_0,b_1
\ldots b_N}(u)(\c_{a_0})_{\d}^{\a}.
\ee
In general by this relation  the spinorial $RTT$ algebra, the generators of
which are contained in the matrix $T$,  is mapped to the
ordinary $RTT$ algebra, the generators of which are matrix elements of
$\mathbb{T}$.  Moreover, it provides a way to solve the spectral problem
for the trace of the ordinary monodromy matrix by solving the spectral
problem for a trace involving the spinorial monodromy matrix.

The entries of the inverse-transpose $\hat{\mathbb{T}}_{ij}(u)=
{\mathbb{T}}^{-1}_{N+1-j,N+1-i}(u)$ to the monodromy
matrix ${\mathbb{T}}$ defined over the fundamental (vector) auxiliary
space used in \cite{lprs} is given by the quantum minors
divided by the quantum determinant. In contrast, 
due to the inversion relation for $L$ (\ref{invL}) the  
inverse of spinor monodromy matrix $T(u)$
 is given
by the same matrix with the shifted  spectral parameter.

\section{Even-dimensional orthogonal algebras}
\setcounter{equation}{0}

In  sect. 3 we have obtained  the spinorial 
$\check{\mathfrak{R}}$-matrix in form of the Euler Beta-function for an
arbitrary orthogonal algebra. Now we shall consider  low rank
examples corresponding to the $D$ series. The universal expression
(\ref{sons})
results in explicit forms using the corresponding
characteristic polynomial in the invariant $z$
and the resulting spectral decomposition.
Details about the invariant $z$ and relations following from 
its characteristic polynomial are considered in the Appendix.

\subsection{The $so(4)$ case}

In this case we have the characteristic polynomial
\be\lb{w4}
W_4=z(z^2-1)(z^2-4)=0.
\ee
In other words, in the case  $d=4$ any function of $z$ is represented by a
polynomial of fourth degree. The roots of $W_4$ are the eigenvalues of $z$
and we are lead to the spectral decomposition
\be\lb{ro44}
\check{\mathfrak{R}}^{so(4)}(z|u)=B\big(\frac{z+1-u}2,u\big)=\sum_{k=-2}
^2B\big(\frac{k+1-u}2,u\big)P_k,
\ee
Here the sum goes over the roots of $W_4$: $z_k=0,\pm1,\pm2$ and $P_k$
are projection operators on the  corresponding eigenspace,
\be\lb{po4}
P_0=\frac14(z^2-1)(z^2-4),,\qquad P_{\pm1}=-\frac1{3!}(z\pm
1)(z^2-4),\qquad P_{\pm2}=\frac1{4!}(z^2-1)(z\pm2).
\ee
Using (\ref{w4}) one can check the properties
$
P_kP_\ell=P_k \d_{k\ell},\ \sum_{k=-2}^2P_k=1
$.
Further, using the functional equation
$
B(x+1,y)=\frac x{x+y}B(x,y),
$
one deduces the explicit form of the spinorial $R$ matrix
$$
\check{\mathfrak{R}}^{so(4)}(z|u)=B\big(\frac12-\frac u2,u\big)
\Big(\frac{1-u}{1+u}(P_2+P_{-2})
+P_0\Big)+B\big(-\frac u2,u\big)\Big(P_1-P_{-1}\Big) =
$$ 
$$ \check{\mathfrak{R}}^{(1)}_{so(4)}(u|z) + 
\check{\mathfrak{R}}^{(2)}_{so(4)}(u|z).
$$
It separates  into two parts  given by even and odd functions of $z$, 
respectively. Both parts satisfy the $ RRR $ Yang-Baxter relation separately.

$z$ is acting on $\mathcal{S}_1 \otimes \mathcal{S}_2 $ and the permutation
operator is given by
\be\lb{pe4}
\mathcal{P}_{12}=P_0+P_1-P_{-1}-P_2-P_{-2}=\frac16(z^4-2z^3-7z^2+8z+6)=
\ee
$$
=\frac14\Big({\mathbb{I}}_{12}
+\frac1{(1!)^2}\c^a_1\c^a_2-\frac1{(2!)^2}\c^{ab}_1\c^{ab}_2-\frac1{(3!)^2}
\c^{abc}_1\c^{abc}_2+\frac1{(4!)^2}\c^{abcd}_1\c^{abcd}_2\Big).
$$

Consider now both parts of the spinorial $R$ matrix in more detail. 
The simpler part
$$
\check{\mathfrak{R}}^{(1)}_{so(4)}(u|z)=-P_1+P_{-1}=-\frac z3(z^2-4)
=-\frac1{144}\c^{abc}_1c^{abc}_2+\frac34\c^a_1\c^a_2,
$$ 
is  given by an odd function of $z$. The corresponding $R$-matrix without
check,
${\mathfrak{R}}^{(1)}=\mathcal{P}_{12}\check{\mathfrak{R}}^{(1)}$, is 
even in $z$,
 $$
{\mathfrak{R}}_{so(4)}^{(1)}(u|z)
=-P_1-P_{-1}=-\frac{z^2(z^2-4)}3=\frac12(1+\c^5_1\c^5_2),
\qquad \c^5=\c^1\c^2\c^3\c^4,
$$
and is trivial, i.e. diagonal and independent of the spectral parameter, with 
the following non-vanishing entries:
\be\lb{r411}
{\mathfrak{R}}^{12}_{12}={\mathfrak
{R}}^{13}_{13}={\mathfrak{R}}^{21}_{21}={\mathfrak{R}}^{24}_{24}
={\mathfrak{R}}^{31}_{31}={\mathfrak{R}}^{34}_{34}={\mathfrak{R}}
^{42}_{42}={\mathfrak{R}}^{43}_{43}=1. 
\ee
 The RLL Yang-Baxter relation 
$$
{\mathfrak{R}}_{so(4)}^{(1)}L_1^{ab}(u)L_1^{bc}(v)=
L_2^{ab}(v)L_1^{bc}(u){\mathfrak{R}}_{so(4)}^{(1)}
$$
reduces here to the identity
$$
(1+\c^5_1\c^5_2)\c_1^{ab}\c_2^{bc}=\c_2^{ab}\c_1^{bc}
(1+\c^5_1\c^5_2).
$$
We have also the Yang-Baxter RRR relation in the trivial form
$$
(1+\c^5_1\c^5_2)(1+\c^5_1\c^5_3)(1+\c^5_2\c^5_3)=
(1+\c^5_2\c^5_3)(1+\c^5_1\c^5_3)(1+\c^5_1\c^5_2).
$$ 

The part of the spinorial spinorial $R$ matrix  even  in $z$
$$
\check{\mathfrak{R}}_{so(4)}^{(2)}(u|z)=\frac{u+2}4(1+\c^5_1\c^5_2)
-\frac u{16}\c^{ab}_1\c^{ab}_2,
$$ 
corresponds to the $R$-matrix without check  also  even in
 $z$, 
 $$
{\mathfrak{R}}_{so(4)}^{(2)}(u|z)
=(u-1)(P_2+P_{-2})+(1+u)P_0=\frac{u-1}{12}z^2(z^2-1)
+\frac{u+1}4(z^2-1)(z^2-4)=
$$
$$
=(u+\frac12)\frac13(z^4-4z^2+3)-\frac12(z^2-1)=\frac{2u+1}
4(1+\c^5_1\c^5_2)-\frac1{16}\c_1^{ab}\c_2^{ab},
$$
The latter is represented by the matrix with the  following  non-vanishing entries:
$$
{\mathfrak{R}}^{11}_{11}(u)={\mathfrak{R}}^{22}_{22}(u)={\mathfrak{R}}
^{33}_{33}(u)={\mathfrak{R}}^{44}_{44}(u)=u+1,
$$
\be\lb{ro4}
{\mathfrak{R}}^{14}
_{14}(u)={\mathfrak{R}}^{23}_{23}(u)={\mathfrak{R}}^{32}_{32}(u)
={\mathfrak{R}}^{41}_{41}(u)=u,
\ee
$$
{\mathfrak{R}}^{14}_{41}(u)={\mathfrak{R}}^{23}_{32}(u)=1=
{\mathfrak{R}}^{32}_{23}(u)={\mathfrak{R}}^{41}_{14}(u).
$$

Thus this matrix has the block form: its  
entries ${\mathfrak{R}}^{\a\b}_{\c\d}(u)$ differ from zero only
if $\a$, $\b$, $\c$ and $\d$ belong to sets $(1,4)$ or ($2,3)$.

It is important to notice that the two different parts
$\check{\mathfrak{R}}^{(1)}_{so(4)}(u)$ and $\check{\mathfrak{R}}
^{(2)}_{so(4)}(u)$, which are odd and even functions of $z$
correspondingly, are distinguished by   {\it{chirality}}. 
Indeed these
solutions are proportional to the chiral projectors $\Pi_+$ and $\Pi_-$,
respectively, where
$$ \Pi_{\pm} = \frac12(1 \pm \c_1^5\c_2^5). $$ 
One deduces from (\ref{imz}) the useful formulae
$$
\Pi_+ = -\frac13z^2(z^2-4),\qquad
\qquad \Pi_- = \frac13(z^2-1)(z^2-3).
$$
 This chiral property ensures the 
{\it{consistency}} of these solutions: both of them intertwine a pair of
$L$-operators i.e. obey the $RLL$-relation linear in $R$, but also satisfy
the trilinear $RRR$-relation not only separately, but also in arbitrary 
combination, due to their orthogonality.

The chiral projectors 
$\Pi_\pm$ separate the  $16$-dimensional representation space $\mathcal{S}_1
\otimes \mathcal{S}_2$
into two  eight-dimensional chiral subspaces. In particular, the
subspace corresponding to $\Pi_+$ is spanned by eight eigenvectors 
of $z$ corresponding to the eigenvalues $\pm1$, while the six eigenvectors,
corresponding to the zero eigenvalue of $z$ as well as vectors corresponding 
to the eigenvalues $\pm2$ span the other chiral subspace.
 In terms of the projectors of the eigenspaces this
reads as
$$
\Pi_+P_{\pm1}=P_{\pm1}\Pi_+=P_{\pm1},\qquad \Pi_-P_{\pm1}=
0=P_{\pm1}\Pi_-,
$$
$$
\Pi_-P_0=P_0\Pi_-=P_0,
\qquad \Pi_-P_{\pm2}=P_{\pm2}\Pi_-=P_{\pm2},
$$
$$
\Pi_+P_0=0=P_0\Pi_+,\qquad \Pi_+P_{\pm2}=0=P_{\pm2}\Pi_+.
$$

Note that due to the definite chirality of 
${\mathfrak{R}}^{(1)}_{so(4)}(u)$ and
${\mathfrak{R}}^{(2)}_{so(4)}(u)$ the expressions like $tr_2\big({\mathfrak
{R}}^{(1)}_{so(4)}(u)\c^a_2{\mathfrak{R}}^{(1)}_{so(4)}(v)\c^b_2\big)$ and
$tr_2\big({\mathfrak{R}}^{(2)}_{so(4)}(u)\c^a_2{\mathfrak{R}}^{(2)}_{so(4)}
(v)\c^b_2\big)$ vanish, but the non-diagonal expression 
$
tr_2\big({\mathfrak{R}}^{(1)}_{so(4)}(u)\c^a_2{\mathfrak{R}}^{(2)}_{so
(4)}(v)\c^b_2\big) $ does not vanish and at $2v+1=u$ results in the 
fusion to the $L$ matrix
(\ref{losc}), 
$$
tr_2\big({\mathfrak{R}}^{(1)}_{so(4)}(2v+1)\c^a_2{\mathfrak{R}}^{(2)}
_{so(4)}(v)\c^b_2\big)=(2v+1)\d^{ab}-\frac12c_1^{ab} = L_1^{ab}(2v+1). 
$$

\subsection{ABA for the $so(4)$ case}

The observed  simple structure of the spinorial ${\mathfrak{R}}^{so(4)}$
matrix is helpful for the diagonalization of  the trace of the spinorial monodromy matrix
\be\lb{to4}
T^{(2)}(u)={\mathfrak{R}}^{(2)}_{01}(u){\mathfrak{R}}^{(2)}_{02}(u)
\ldots {\mathfrak{R}}^{(2)}_{0N}(u),
\ee
or in components
$$
T^{\a_0,\a_1\ldots\a_N}_{\b_0,\b_1\ldots\b_N}(u)={\mathfrak{R}}
^{\a_0\a_1}_{\c_0\b_1}(u)\ldots{\mathfrak{R}}^{\d_0\a_N}_{\b_0\b_N}(u).
$$

 The explicit form of the matrix ${\mathfrak{R}}^{(2)}_{0k}(u)$
(\ref{ro4}) results in the the following representation as a $4\times4$ matrix in
the
auxiliary space $\mathcal{S}_0$ with operator valued elements
acting in the quantum space $\mathcal{S}_k$
\be\lb{ro440k}
{\mathfrak{R}}^{(2)}_{0k}(u)=\left(\ba{cccc}{\mathfrak{R}}^1_1(u)&0&0&
{\mathfrak{R}}^1_4(u)\\0&{\mathfrak{R}}^2_2(u)&{\mathfrak{R}}^2_3(u)&0\\
0&{\mathfrak{R}}^3_2(u)&{\mathfrak{R}}^3_3(u)&0\\ {\mathfrak{R}}^4_1(u)&
0&0&{\mathfrak{R}}^4_4(u)u\ea\right).
\ee
Its diagonal $4\times 4$ matrix components are diagonal
$$
{\mathfrak{R}}^1_1(u)=diag(u+1, 0,0,u), \ \  
{\mathfrak{R}}^2_2(u)=diag(0,u+1,u,0),
$$
$$
{\mathfrak{R}}^3_3(u)=diag(0, u,u+1,0,\qquad
{\mathfrak{R}}^4_4(u)=diag(u,0,0,u+1),
$$
and have to be regarded as the elements of the Cartan subalgebra.
The off-diagonal elements have to be  regarded as lowering 
$$
{\mathfrak{R}}^1_4(u)=\mathbf{e}_{4 1},  
\qquad{\mathfrak{R}}^2_3(u)=\mathbf{e}_{32},
$$
and rising
$$
{\mathfrak{R}}^3_2(u)=\mathbf{e}_{2 3}
,\qquad{\mathfrak{R}}^4_1(u)=\mathbf{e}_{14},
$$
generators correspondingly. 
Consequently, the monodromy matrix (\ref{to4}) defined as an ordered 
matrix product of the factors (\ref{ro44}) preserves the block form,
 \be\lb{to41}
{{T}}^{(2)}(u)=\left(\ba{cccc}{{T}}^1_1(u)&0&0&
{{T}}^1_4(u)\\0&{{T}}^2_2(u)&{{T}}^2_3(u)&0\\
0&{{T}}^3_2(u)&{{T}}^3_3(u)&0\\ {{T}}^4_1(u)&
0&0&{{T}}^4_4(u)u\ea\right),
\ee
with Cartan elements ${{T}}^\a_\a(u)$ and  rising 
$C_1(u)={{T}}^4_1(u)$, $C_2(u)={{T}}^3_2(u)$ and
lowering $B_2(u)={{T}}^2_3(u)$, $B_1(u)={{T}}^1_4(u)$ generators. 
We have obtained a representation of the spinorial $RTT$ algebra 
and its decomposition  
into a representation of two subalgebras of the $s\ell(2)$ type Yangian.

This becomes more evident by a similarity transformation with
the $4 \times 4$ matrix $V$
\be\lb{v}
V= \mathbf{e}_{11} + \mathbf{e}_{24} + \mathbf{e}_{33} + \mathbf{e}_{42} 
,\qquad\qquad V^{-1}=V.
\ee
 One  calculates easily
\be\lb{vtv}
V{{T}}^{(2)}(u)V=\left(\ba{cccc}
{{T}}^1_1(u)&{{T}}^1_4(u)&0&0\\
{{T}}^4_1(u)&{{T}}^4_4(u)&0&0\\
0&0&{{T}}^3_3(u)&{{T}}^3_2(u)\\ 
0&0&{{T}}^2_3(u)&{{T}}^2_2(u)\ea\right),
\ee
and
\be\lb{vrv}
(V\otimes V){\mathfrak{R}}^{(2)}(u)(V\otimes V)=
\left(\ba{cc}{\mathfrak{R}}^{I}&0\\0&{\mathfrak{R}}^{II}
\ea\right),
\ee
with the $8\times8$ block-matrices
$$
{\mathfrak{R}}^{I}(u)=\left(\ba{cccccccc}
{\mathfrak{R}}^{11}_{11}(u)&0&0&0&0&0&0&0\\
0&{\mathfrak{R}}^{14}_{14}(u)&0&0&{\mathfrak{R}}^{14}_{41}(u)&0&0&0\\
0&0&0&0&0&0&0&0\\
0&0&0&0&0&0&0&0\\
0&{\mathfrak{R}}^{41}_{14}(u)&0&0&{\mathfrak{R}}^{41}_{41}(u)&0&0&0\\
0&0&0&0&0&{\mathfrak{R}}^{44}_{44}(u)&0&0\\
0&0&0&0&0&0&0&0\\
0&0&0&0&0&0&0&0
\ea\right),
$$
$$
{\mathfrak{R}}^{II}(u)=\left(\ba{cccccccc}
0&0&0&0&0&0&0&0\\
0&0&0&0&0&0&0&0\\
0&0&{\mathfrak{R}}^{33}_{33}(u)&0&0&0&0&0\\
0&0&0&{\mathfrak{R}}^{32}_{32}(u)&0&0&
{\mathfrak{R}}^{32}_{23}(u)&0\\
0&0&0&0&0&0&0&0\\
0&0&0&0&0&0&0&0\\
0&0&0&{\mathfrak{R}}_{32}^{23}(u)&0&0&
{\mathfrak{R}}^{23}_{23}(u)&0\\ 
0&0&0&0&0&0&0&
{\mathfrak{R}}^{22}_{22}(u)
\ea\right).
$$
Disregarding in these blocks the lines and columns with zero entries, we
recognize in each of them the fundamental $R$ matrix of $s\ell(2)$.

Consider now the general RTT relation substituting for $T$ instead of the particular
monodromy matrix (\ref{to4}) a generic algebra valued  $4\times4$ matrix.
Whereas in the  $s\ell(4)$ case the corresponding relation results in the
first step in 256 non-trivial relations, here the relation with the $so(4)$ spinorial
$R$ matrix  (\ref{ro4}) leads to 64 non-trivial relations from the matrix
elements with the
indices belonging to sets $(1,4)$ and $(2,3)$,
 to further relations  of the form $0=0$ and , most interesting, to 
such  with non-trivial r.h.s. and  zero l.h.s
or vice versa.  
 For example,
\be\lb{rtt4}
{\mathfrak{R}}^{\a_1\a_2}_{\b_1\b_2}(u-v)T^{b_1}_{\c_1}(u)T^{b_2}_{\c_2}(v)=
T^{a_2}_{\b_2}(v)T^{a_1}_{\b_1}(u){\mathfrak{R}}^{\b_1\b_2}_{\c_1\c_2}(u-v),
\ee
for $(\a_1,\a_2)=(1,1)$, $(\c_1,\c_2)=(1,2)$ gives
$$
T^1_1(u)T^1_2(v)=0.
$$
Combining all these relations one deduces, that the general $so(4)$ spinor
monodromy matrix  has the form (\ref{to41}), i.e. 
$$
T^1_2(u)=T^1_3(u)=T^1_4(u)=T^2_4(u)=T^3_4(u)=0,
$$
and
$$
T^2_1(u)=T^3_1(u)=T^4_1(u)=T^4_2(u)=T^4_3(u)=0.
$$
Thus also in the case of  arbitrary representations  in the quantum space the
$so(4)$ monodromy matrix generates an algebra equivalent to two 
independent $s\ell(2)$ Yangian algebras and the spectral problem for its
trace leads to the ordinary $s\ell(2)$ ABA. 

In particular,  this applies to the spinor-vector monodromy matrix (\ref{tv}).
Recall, that it produces the $so(4)$ vector-vector monodromy matrix by
fusion (\ref{t2t}), 
\be\lb{ttt4}
{\mathbb{T}}^{a_0,\,\,a_1,\ldots a_N}_{a_0,\,\,b_1,\ldots b_N}(u)=\frac14
tr\Big( T^{a_1,\ldots a_N}_{c_1,\ldots c_N}(u+\frac12)\c^{a_0}
{T^{-1}}^{c_1,\ldots c_N}_{b_1,\ldots b_N}(u+\frac32)\c_{a_0}\Big).
\ee
We have to substitute $\b=1$ for $so(4)$. 

 Taking into account that $T$ and $\bar T=
T^{-1}$ have the form (\ref{to41}), one can calculate
explicitly the matrix elements of the vector monodromy in terms of products
of the matrix elements  of the spinor mondromy. In particular for the trace
we obtain
\be\lb{ttt3}
{\mathbb{T}}^{a_0}_{a_0}(u)=\frac12\Big((T^1_1+T^4_4)(\bar T^2_2+
\bar T^3_3)+(T^2_2+T^3_3)(\bar T^1_1+\bar T^4_4)\Big),
\ee
where we have abbreviated $T^\a_\b(u+\frac12)=\Big(L_1(u+\frac12)\ldots L_N(u+
\frac12)\Big)^\a_\b$ and $\bar T=T^{-1}(u+\frac32)$.
The vector-vector $so(4)$ monodromy matrix is
given by  sums of products of the  spinor-vector transfer
matrix elements which obey  $s\ell(2)$ type Yangian relations.

At the end of this subsection we consider the $RTT$ algebra generated by  
${\mathfrak{R}}^{(1)}_{so(4)}(u)$. Due to its chirality and the diagonal
character it leads to a trivial solution for $T(u)$ of the RTT-relation.  For 
instance, $(\a_1,\a_2)=(1,1)$, $(\c_1,\c_2)=(1,2)$ and $(\c_1,\c_2)
=(2,1)$ by (\ref{r411}) leads to 
$$
T^1_1(v)T^1_2(u)=0=T^1_2(v)T^1_1(u).
$$ 
The remaining 
RTT-relations imply that the most general monodromy matrix $T(u)$
intertwined by ${\mathfrak{R}}^{(1)}_{so(4)}(u)$ is given by the diagonal
matrix:
$$
T(u)= diag\big( T^1_1(u),T^2_2(u),T^3_3(u),T^4_4(u)\big) 
$$
which has vanishing rising and lowering generators.

We summarize the above results:

\begin{proposition}

The spinorial $R$ matrix with $so(4)$ symmetry 
${\mathfrak{R}}^{(2)}$ (acting in the chiral subspace
 of the $\Pi_- $ projection) generates the spinorial $RTT$ algebra
decomposing into two subalgebras of the $s\ell (2)$ Yangian type. 
The spinorial $RTT$ algebra generated by  ${\mathfrak{R}}^{(1)}$ (acting in
the chiral subspace of the  $\Pi_+ $ projection) is a trivial commuting
algebra. 
In this way, the ordinary $s\ell_2$ ABA allows to construct  
solutions of the ABA of for the spinorial Yangian of $so(4)$ type.

\end{proposition}

\subsection{The $so(6)$ case}
In this case $z$ obeys
\be \label{w6}
 W_6= z(z^2-1)(z^2-4)(z^2-9)=0, \ee
has seven different eigenvalues: $0,\pm1,\pm2,\pm3$
and the projectors on the corresponding  eigenspaces are given by
\be\lb{pp6}
P_{\pm3}=\frac{z(z^2-1)(z^2-4)(z\pm3)}{6!},\quad
P_{\pm2}=\frac{z(z^2-1)(z^2-9)(z\pm2)}{5!},\quad
\ee
$$
\!\!\!\!\!\!\!\!\!\!\!\!\!\!\!\!\!\!\!\!P_{\pm1}=\frac{z(z^2-4)(z^2-9)(z\pm1)}
{2\cdot4!},\quad P_0=-\frac{(z^2-1)(z^2-4)(z^2-9)}{3!}.
$$
Further, the permutation operator has the spectral expansion
\be\lb{perm6}
\mathcal{P}_{12}=P_0+P_1-P_{-1}-P_2-P_{-2}-P_3+P_{-3}.
\ee
Again, the $R$-operator  decomposes  into even and odd parts in $z$
$$
\check{\mathfrak{R}}_{so(6)}(z|u)=B\big(\frac{z+1-u}2,u\big)=\frac{B
\big(-\frac12-\frac u2,u\big)}{u-1}\check{\mathfrak{R}}^{(1)}_{so(6)}+
\frac{B\big(-1-\frac u2,u\big)}{u-2}\check{\mathfrak{R}}^{(2)}_{so(6)}=
$$
\be\lb{cR6}
=\frac{B\big(-\frac12-\frac u2,u\big)}{u-1}\Big((u-1)(P_2+P_{-2})-
({u+1})P_0\Big)+
\ee
$$
+\frac{B\big(-1-\frac u2,u\big)}{u-2}\Big((u+2)
(P_1-P_{-1})+({u-2})(P_{-3}-P_3)\Big).
$$
According to (\ref{perm6}) the $R$-matrices without check are
\be\lb{r61}
{\mathfrak{R}}^{(1)}_{so(6)}(z|u)=-(u-1)(P_2+P_{-2})-
({u+1})P_0,
\ee
and
\be\lb{r62}
{\mathfrak{R}}^{(2)}_{so(6)}(z|u)=(u+2)
(P_1+P_{-1})+({u-2})(P_{-3}+P_3).
\ee
We introduce in analogy to the $so(4)$ case
 $\c^7_1\c^7_2=\frac4{45}(z^6-\frac{25}2
z^4+34z^2-\frac{45}4)$ 
and observe that the two parts of the spinorial $R$
matrix are
proportional to the chiral  projectors $\Pi_{\mp}$, respectively, where 
$$
\Pi_\pm=\frac12(1\pm\c^7_1\c^7_2)=\frac12\pm\frac2{45}\Big(
z^6-\frac{25}2z^4+34z^2-\frac{45}4\Big).
$$  
Using the characteristic polynomial $W_6$ (\ref{w6}) 
 one can prove easily the  relations
$
P_iP_j=\d_{ij}P_i, \sum_{i=-3}^3P_i=1,
$
as well as
\be\lb{pr61}
\Pi_+P_{\pm(2k+1)}=P_{\pm(2k+1)}=P_{\pm(2k+1)}\Pi_+,\quad
\Pi_-P_{\pm(2k+1)}=0=P_{\pm(2k+1)}\Pi_-,
\ee
\be\lb{pr62}
\Pi_-P_{\pm2k}=P_{\pm2k}=P_{\pm2k}\Pi_-,\qquad
\Pi_+P_{\pm2k}=0=P_{\pm2k}\Pi_+,\qquad k=0,1.
\ee

\subsection{ABA for the $so(6)$ case}
In this case we again deal with two spinor-spinor $R$-matrices (\ref{r61})
and (\ref{r62}) acting in the two chiral subspaces. 

\subsubsection{ ${\mathfrak{R}}^{(1)}$, the $\Pi_-$ part}

According to (\ref{r61}) this $R$-matrix
 has the following non-vanishing entries:
$$
{\mathfrak{R}}^{11}_{11}(u)={\mathfrak{R}}^{22}_{22}(u)=
{\mathfrak{R}}^{33}_{33}(u)={\mathfrak{R}}^{44}_{44}(u)=
{\mathfrak{R}}^{55}_{55}(u)={\mathfrak{R}}^{66}_{66}(u)=
{\mathfrak{R}}^{77}_{77}(u)={\mathfrak{R}}^{88}_{88}(u)=-6(u+1),
$$
$$
{\mathfrak{R}}^{14}_{14}(u)={\mathfrak{R}}^{16}_{16}(u)=
{\mathfrak{R}}^{17}_{17}(u)={\mathfrak{R}}^{23}_{23}(u)=
{\mathfrak{R}}^{25}_{25}(u)={\mathfrak{R}}^{28}_{28}(u)=
{\mathfrak{R}}^{32}_{32}(u)={\mathfrak{R}}^{35}_{35}(u)=
$$
$$
{\mathfrak{R}}^{38}_{38}(u)={\mathfrak{R}}^{41}_{41}(u)=
{\mathfrak{R}}^{46}_{46}(u)={\mathfrak{R}}^{47}_{47}(u)=
{\mathfrak{R}}^{52}_{52}(u)={\mathfrak{R}}^{53}_{53}(u)=
{\mathfrak{R}}^{58}_{58}(u)={\mathfrak{R}}^{61}_{61}(u)=
$$
$$
{\mathfrak{R}}^{64}_{64}(u)={\mathfrak{R}}^{67}_{67}(u)=
{\mathfrak{R}}^{71}_{71}(u)={\mathfrak{R}}^{74}_{74}(u)=
{\mathfrak{R}}^{76}_{76}(u)={\mathfrak{R}}^{82}_{82}(u)=
{\mathfrak{R}}^{83}_{83}(u)={\mathfrak{R}}^{85}_{85}(u)=
-\frac12(5u+7),
$$
\be\lb{r611}
{\mathfrak{R}}^{14}_{41}(u)={\mathfrak{R}}^{16}_{61}(u)=
{\mathfrak{R}}^{17}_{71}(u)={\mathfrak{R}}^{23}_{32}(u)=
{\mathfrak{R}}^{25}_{52}(u)={\mathfrak{R}}^{28}_{82}(u)=
{\mathfrak{R}}^{32}_{23}(u)={\mathfrak{R}}^{35}_{53}(u)=
\ee
$$
{\mathfrak{R}}^{38}_{83}(u)={\mathfrak{R}}^{41}_{14}(u)=
{\mathfrak{R}}^{46}_{64}(u)={\mathfrak{R}}^{47}_{74}(u)=
{\mathfrak{R}}^{52}_{25}(u)={\mathfrak{R}}^{53}_{35}(u)=
{\mathfrak{R}}^{58}_{85}(u)={\mathfrak{R}}^{61}_{16}(u)=
$$
$$
{\mathfrak{R}}^{64}_{46}(u)={\mathfrak{R}}^{67}_{76}(u)=
{\mathfrak{R}}^{71}_{17}(u)={\mathfrak{R}}^{74}_{47}(u)=
{\mathfrak{R}}^{76}_{67}(u)={\mathfrak{R}}^{82}_{28}(u)=
{\mathfrak{R}}^{83}_{38}(u)={\mathfrak{R}}^{85}_{58}(u)=
-\frac12(7u+5).
$$
In analogy to the  $so(4)$ case, considering $(\a_1,\a_2)=(1,1)$ and
$(\c_1,\c_2)=(1,2),\,\,(1,3),$ $(1,5),\,\,(1,8)$, in (\ref{rtt4}), one
obtains for the arbitrary $8 \times 8 $ monodromy matrix $T(u)$:
$$
T^1_1(u)T^1_2(v)=0,\qquad T^1_1(u)T^1_3(v)=0,\qquad
T^1_1(u)T^1_5(v)=0,\qquad T^1_1(u)T^1_8(v)=0,
$$
 which have the solution
$$
T^1_2(v)= T^1_3(v)=T^1_5(v)= T^1_8(v)=0.
$$
Similarly, considering $(\a_1,\a_2)=(2,2)$ and $(\c_1,\c_2)=(2,1),\,
(2,4),\,(2,6),\,(2,7)$ leads to
$$
T^2_1(v)= T^2_4(v)=T^2_6(v)= T^2_7(v)=0,
$$
and further to
$$
T^3_1(v)=0 T^3_4(v)=T^3_6(v)=0 T^3_7(v)=0, \ \ \
T^4_2(v)= T^4_3(v)=T^4_5(v)= T^4_8(v)=0,
$$
$$
T^5_1(v)= T^5_4(v)=T^5_6(v)= T^5_7(v)=0, \ \ \ 
T^6_2(v)= T^6_3(v)=T^6_5(v)= T^6_8(v)=0,
$$ $$
T^7_2(v)= T^7_3(v)=T^7_5(v)=T^7_8(v)=0, \ \ \
T^8_1(v)= T^8_4(v)= T^8_6(v)= T^8_7(v)=0.
$$
Thus the general monodromy matrix has the block-form:
\be\lb{mm6}
T(u)=\left(\ba{cccccccc}T^1_1(u)&0&0&T^1_4(u)&0&T^1_6(u)&T^1_7(u)&0\\
0&T^2_2(u)&T^2_3(u)&0&T^2_5(u)&0&0&T^2_8(u)\\
0&T^3_2(u)&T^3_3(u)&0&T^3_5(u)&0&0&T^3_8(u)\\
T^4_1(u)&0&0&T^4_4(u)&0&T^4_6(u)&T^4_7(u)&0\\
0&T^5_2(u)&T^5_3(u)&0&T^5_5(u)&0&0&T^5_8(u)\\
T^6_1(u)&0&0&T^6_4(u)&0&T^6_6(u)&T^6_7(u)&0\\
T^7_1(u)&0&0&T^7_4(u)&0&T^7_6(u)&T^7_7(u)&0\\
0&T^8_2(u)&T^8_3(u)&0&T^8_5(u)&0&0&T^8_8(u)
\ea\right) .
\ee
By renaming indices or interchanging rows and columns this matrix
can be turned to the block-diagonal form, which means that the $8\times8$
matrix elements of spinor $RTT$ algebra generators
reduce to two independent subsets each generating  a  $s\ell(4)$ type RTT algebra.
This becomes  evident
by the similarity transformation with the matrix $V$:
\be\lb{v6}
V= \mathbf{e}_{11} +  \mathbf{e}_{27} +  \mathbf{e}_{36} + 
 \mathbf{e}_{44} +  \mathbf{e}_{55} +  \mathbf{e}_{63} + 
 \mathbf{e}_{72} +  \mathbf{e}_{88}. 
\ee
One obtains for the monodromy matrix (\ref{mm6}):
\be\lb{vmv6}\!\!\!\!
VT(u)V=\left(\ba{cccccccc}
T^1_1(u)&T^1_7(u)&T^1_6(u)&T^1_4(u)&0&0&0&0\\
T^7_1(u)&T^7_7(u)&T^7_6(u)&T^7_4(u)&0&0&0&0\\
T^6_1(u)&T^6_7(u)&T^6_6(u)&T^6_4(u)&0&0&0&0\\
T^4_1(u)&T^4_7(u)&T^4_6(u)&T^4_4(u)&0&0&0&0\\
0&0&0&0&T^5_5(u)&T^5_3(u)&T^5_2(u)&T^5_8(u)\\
0&0&0&0&T^3_5(u)&T^3_3(u)&T^3_2(u)&T^3_8(u)\\
0&0&0&0&T^2_5(u)&T^2_3(u)&T^2_2(u)&T^2_8(u)\\
0&0&0&0&T^8_5(u)&T^8_3(u)&T^8_2(u)&T^8_8(u)
\ea\right).
\ee
The $R$-matrix ${\mathfrak{R}}^{(1)}_{so(6)}(u)$ is also 
transformed by this similarity transformation to the 
block-diagonal form similar to (\ref{vrv}). 
We obtain two blocks of the $s\ell(4)$ $R$ matrix form.

\subsubsection{${\mathfrak{R}}^{(2)}$, the $\Pi_+$ part }

Consider now the second part acting in the chiral sector $\Pi_+$,
$$
{\mathfrak{R}}^{(2)}_{so(6)}(u)=\frac{u+2}{24}z^2(z^2-4)(z^2-9)
+\frac{u-2}{360}z^2(z^2-1)(z^2-4).
$$
It appears as a matrix with the following non-vanishing entries: 
$$\!\!\!\!\!\!\!\!\!\!
{\mathfrak{R}}^{18}_{18}(u)={\mathfrak{R}}^{27}_{27}(u)=
{\mathfrak{R}}^{36}_{36}(u)={\mathfrak{R}}^{45}_{45}(u)=
{\mathfrak{R}}^{54}_{54}(u)={\mathfrak{R}}^{63}_{63}(u)=
{\mathfrak{R}}^{72}_{72}(u)={\mathfrak{R}}^{81}_{81}(u)=u-1,
$$
$$\!\!\!\!\!\!\!\!\!\!\!\!\!\!\!\!\!
{\mathfrak{R}}^{12}_{12}(u)={\mathfrak{R}}^{13}_{13}(u)=
{\mathfrak{R}}^{15}_{15}(u)={\mathfrak{R}}^{21}_{21}(u)=
{\mathfrak{R}}^{24}_{24}(u)={\mathfrak{R}}^{26}_{26}(u)=
{\mathfrak{R}}^{31}_{31}(u)={\mathfrak{R}}^{34}_{34}(u)=
$$
$$\!\!\!\!\!\!\!\!\!\!
{\mathfrak{R}}^{37}_{37}(u)={\mathfrak{R}}^{42}_{42}(u)=
{\mathfrak{R}}^{43}_{43}(u)={\mathfrak{R}}^{48}_{48}(u)=
{\mathfrak{R}}^{51}_{51}(u)={\mathfrak{R}}^{56}_{56}(u)=
{\mathfrak{R}}^{57}_{57}(u)={\mathfrak{R}}^{62}_{62}(u)=
$$
$$\!\!\!\!\!\!\!\!\!\!
{\mathfrak{R}}^{65}_{65}(u)={\mathfrak{R}}^{68}_{68}(u)=
{\mathfrak{R}}^{73}_{73}(u)={\mathfrak{R}}^{75}_{75}(u)=
{\mathfrak{R}}^{78}_{78}(u)={\mathfrak{R}}^{84}_{84}(u)=
{\mathfrak{R}}^{86}_{86}(u)={\mathfrak{R}}^{87}_{87}(u)=u+2,
$$
\be\lb{r621}\!\!\!\!\!
{\mathfrak{R}}^{18}_{45}(u)={\mathfrak{R}}^{27}_{36}(u)=
{\mathfrak{R}}^{45}_{63}(u)={\mathfrak{R}}^{36}_{54}(u)=
{\mathfrak{R}}^{63}_{72}(u)={\mathfrak{R}}^{54}_{81}(u)=
{\mathfrak{R}}^{18}_{72}(u)={\mathfrak{R}}^{27}_{81}(u)=
\!\!\!\!
\ee
$$\!\!\!\!\!\!\!\!\!\!\!\!\!\!\!\!\!
{\mathfrak{R}}^{54}_{36}(u)={\mathfrak{R}}^{63}_{45}(u)=
{\mathfrak{R}}^{72}_{18}(u)={\mathfrak{R}}^{81}_{27}(u)=
{\mathfrak{R}}^{45}_{18}(u)={\mathfrak{R}}^{36}_{27}(u)=
{\mathfrak{R}}^{72}_{63}(u)={\mathfrak{R}}^{81}_{54}(u)=1,
$$
$$\!\!\!\!\!\!\!\!\!\!
{\mathfrak{R}}^{81}_{36}(u)={\mathfrak{R}}^{72}_{45}(u)=
{\mathfrak{R}}^{63}_{18}(u)={\mathfrak{R}}^{54}_{27}(u)=
{\mathfrak{R}}^{27}_{54}(u)={\mathfrak{R}}^{18}_{63}(u)=
{\mathfrak{R}}^{45}_{72}(u)={\mathfrak{R}}^{36}_{81}(u)=-1.
$$

Again, the RTT relation (\ref{rtt4}) at index values 
$(\a_1,\a_2)=(1,2),\,\,(1,3),\,\,
(1,5),$ and $(\c_1,\c_2)=(1,1),\,\,(1,4),\,\,(1,6),\,\,(1,7)$ is solved by
$$
T^2_1(v)= T^2_4(v)=T^2_6(v)= T^2_7(v)=0, \ 
T^3_1(v)= T^3_4(v)= T^3_6(v)= T^3_7(v)=0, $$ $$ 
T^5_1(v)= T^5_4(v)=T^5_6(v)= T^5_7(v)=0.
$$
Choosing $(\a_1,\a_2)=(1,1),\,\,(1,4),\,\,(1,6),\,\,(1,7),$ and 
$(\c_1,\c_2)=(1,2),\,\,(1,3),\,\,(1,5),$ one deduces
$$
T_2^1(v)= T_2^4(v)=T_2^6(v)=T_2^7(v)=0, \qquad\qquad
T_3^1(v)= T_3^4(v)= T_3^6(v)= T_3^7(v)=0, 
$$ 
$$ 
T_5^1(v)= T_5^4(v)= T_5^6(v)= T_5^7(v)=0.
$$
Finally, considering say $(\a_1,\a_2)=(2,1),\,\,(2,4),\,\,(2,6)$, 
$(\c_1,\c_2)=(2,8)$ and  $(\a_1,\a_2)=(3,7)$, $(\c_1,\c_2)=(3,8)$ 
one obtains:
$$
T_8^1(v)= T_8^4(v)= T_8^6(v)= T_8^7(v)=0,
$$
and similarly
$$
T^8_1(v)= T^8_4(v)=T^8_6(v)= T^8_7(v)=0.
$$
In this way one deduces that the second chiral part ${\mathfrak{R}}^{(2)}_{so(6)}(u)$
also implies the general form (\ref{mm6}) for the general monodromy matrix.

We summarize the results on the $so(6)$ case:

\begin{proposition}

The spinorial $R$ matrices with $so(6)$ symmetry of both chiral sectors
can be separated into blocks of the fundamental $s\ell(4)$ $R$ matrix form.
In each case the resulting $RTT$ algebras are equivalent to two copies of the
Yangian algebra of  $s\ell(4)$ type. The spectral problem of the
trace of the  $so(6)$ monodromy matrices can be treated on the basis of the
known nested ABA for the  $s\ell(4)$  Yangian.

\end{proposition}

\subsection{The $so(8)$ case}
The characteristic polynomial in this case is given by
\be \label{w8}
W_8=z(z^2-1)(z^2-4)(z^2-9)(z^2-16)=0,
\ee
 and the universal expression (\ref{sons}) for the spinorial $R$ matrix is reduced to
\be\lb{r8}
\check{\mathfrak{R}}_{so(8)}(z|u)=B\big(-1-\frac u2,u\big)\Big(P_{-3}-P_3+
\frac{u+2}{u-2}(P_1-P_{-1})\Big)+
\ee
$$
+B\big(-\frac32-\frac u2,u\big)\Big(P_{-4}+P_4+
\frac{u+3}{u-3}(P_2+P_{-2})+\frac{u+3}{u-3}\frac{u+1}{u-1}P_0\Big).
$$
It decomposes into two parts of opposite chirality, both obeying the
Yang-Baxter relations. 

We observe  that  one of the two chiral parts is simpler:
${\mathfrak{R}}_{so(8)}^{(1)}$ is linear in the spectral parameter
and contains two invariant tensors, while ${\mathfrak{R}}_{so(8)}^{(2)}$
is quadratic in $u$ and contains three tensors.

\section{Odd-dimensional orthogonal algebras}
\setcounter{equation}{0}

In the odd-dimensional cases the characteristic polynomial reduces to
$\tilde  W_d = \prod_k(z-z_k) $ (\ref{Wdt}),  where the product runs over
$m+1$ of the $2m+1$ eigenvalues of $z$. This
leads to the  decomposition of the Euler Beta-function in (\ref{sons}):
\be\lb{bf}
\check{\mathfrak{R}}_{12}(z|u)=B\big(\frac{z+1-u}2,u\big)=\sum_kB\big(
\frac{z_k+1-u}2,u\big)P_k,
\ee
where  the summation goes over the $m+1$ roots $z_k = (-1)^k\frac{2k+1}2$ of 
the characteristic polynomial $\tilde W_d$.

\subsection{The $so(3)$ case}

The charactristic polynomial is
\be \label{Wt3}
 \tilde W_3 = (z+\frac12) (z-\frac32). \ee

We have the decomposition
$$
\check{\mathfrak{R}}^{so(3)}(z|u)=B\big(-\frac14-\frac u2,u\big)\cdot P_{-\frac32}+
B\big(\frac34-\frac u2,u\big)\cdot P_{\frac12}=\frac{B(-\frac14-\frac u2,u)}{1-2u}
\Big(2u(z+\frac12)+1\Big),
$$
where $P_{-\frac32}=-\frac12(z-\frac12)$, $P_{\frac12}=
\frac12(z+\frac32)$.
 Unity and the permutation are given by $1=P_{\frac12}+
P_{-\frac32}$, ${\mathcal{P}}_{12}=P_{\frac12}-P_{-\frac32}$.
Thus
\be\lb{ro3}
{\mathfrak{R}}^{so(3)}(z|u)=2u+z+\frac12=2u\mathbb{I}_{12}+\mathcal{P}_{12}.
\ee
Up to a redefinition of the spectral parameter the spinorial
$so(3)$ $R$ matrix coincides with the fundametal $R$ matrix with $s\ell(2)$
symmetry. 
 
Let us compare this expression also with the fundamental $R$ matrix 
(\ref{rzW}) at $\epsilon=-1$ and 
$n=2$ corresponding to the case $sp(2)$. This  $4\times4$ 
matrix has the  following non-vanishing entries:
$$
{R}^{-1,-1}_{-1,-1}(u)=(u+1)(u+2)={R}^{11}_{11}(u),\qquad
{R}^{-1,1}_{-1,1}(u)=u(u+1)={R}^{1,-1}_{1,-1}(u),
$$
$$
{R}^{-1,1}_{1,-1}(u)=2(u+1)={R}^{1,-1}_{-1,1}(u),
$$
and can be rewritten as
$R(u)=2(u+1)(\frac u2I_{12}+P_{12})$, 
i.e. it coincides 
with the $so(3)$ spinor-spinor $R$-matrix after rescaling $2u\to\frac u2$.

The spinorial $RTT$ relation  for the $so(3)$ monodromy matrix (\ref{tv}) 
coincides with the one  for the vector $s\ell(2)$ mondromy matrix. Indeed, denoting 
\be\lb{t22}
T(u)=\left(\ba{cc}A(u)&B(u)\\C(u)&D(u)\ea\right),
\ee
one obtains that the commutation relations between $A$, $B$, $C$ and $D$
are the same as in $s\ell(2)$ case, because the spinor-spinor $R$-matrix 
(\ref{ro3}) intertwining them, coincide up to $u\to 2u$.

The coincidence of the spinorial $R$ matrix of  $so(3)$ symmetry with the
well known Yang formula for $s\ell(2)$ was a central point in the classical
paper by Reshetikhin \cite{r}. The fusion relation (\ref{t2t}) results in 
expressions of the 9 elements of the vector monodromy matrix $\mathbb{T}$
in terms of the 4 spinorial $RTT$ generators in $T(u)$ for all
representations admitting this relation. The study of \cite{lprs}
resulted in relations among the matrix elements of $\mathbb{T}$, in
particular the three elements in the upper triagle are expressed in terms
of one of them. This confirms that  (\ref{t2t}) establishes  the equivalence
of the spinorial $so(3)$ type $RTT$ algebra and the ordinary 
Yangian of  $s\ell(2)$ type.

\subsection{The $so(5)$ case}

The characteristic polynomial is
\be \label{Wt5}
 \tilde W_5 = (z-\frac12)(z+\frac32)(z-\frac52)=0. \ee 

The spinorial $R$ matrix acting in $\mathcal{S}_1 \otimes \mathcal{S}_2$
 can be written as
$$
\check{\mathfrak{R}}^{so(5)}(z|u)=(u+\frac12)(u-\frac32)
(z-\frac12)(z+\frac32)+
2(u+\frac12)(u+\frac32)(z+\frac32)(z-\frac52)+
$$
$$
+(u-\frac12)
(u+\frac32)(z-\frac12)(z-\frac52)=
(u+1)(u+3)I_{12}+u(u+2)\gamma_1^a\gamma_2^a-\frac{u(u+1)}
2\gamma_1^{ab}\gamma_2^{ab},
$$
The permutation is given by 
$$\mathcal{P}_{12}=P_{\frac12}-P_{-\frac32}
-P_{\frac52}$$ 
with the eigenspace projectors 
$$
P_{\frac12}=-\frac14(z+\frac32)(z-\frac52), \ \ 
P_{-\frac32}=\frac18(z-\frac12)(z-\frac52),
$$
\be\lb{p5}
P_{\frac52}=\frac18(z-\frac12)(z+\frac32).
\ee

Their properties 
$
P_{\frac12}^2=P_{\frac12},
\  P_{-\frac32}^2=P_{-\frac32},
\  P_{\frac52}^2=P_{\frac52},
$ 
as well as 
$
P_{\frac12}+P_{-\frac32}+P_{\frac52}=1,\ 
P_{\frac12}\cdot P_{-\frac32}=
P_{\frac52}\cdot P_{\frac12}=P_{-\frac32}\cdot P_{\frac52}=0,
$ 
are checked by the characteristic polynomial $\tilde W_5$.

The $16\times16$ $R$ matrix $\mathfrak{R} = \mathcal{P}_{12} \check
 {\mathfrak{R}}$
has the following non-vanishing matrix elements:
$$
{\mathfrak{R}}^{11}_{11}(u)={\mathfrak{R}}^{22}_{22}(u)=
{\mathfrak{R}}^{33}_{33}(u)={\mathfrak{R}}^{44}_{44}(u)=
(2u+1)(2u+3),
$$
$$
{\mathfrak{R}}^{12}_{12}(u)={\mathfrak{R}}^{13}_{13}(u)=
{\mathfrak{R}}^{21}_{21}(u)={\mathfrak{R}}^{24}_{24}(u)=
{\mathfrak{R}}^{31}_{31}(u)={\mathfrak{R}}^{34}_{34}(u)=
{\mathfrak{R}}^{42}_{42}(u)={\mathfrak{R}}^{43}_{43}(u)=
2u(2u+3),
$$
$$
{\mathfrak{R}}^{14}_{14}(u)={\mathfrak{R}}^{23}_{23}(u)=
{\mathfrak{R}}^{32}_{32}(u)={\mathfrak{R}}^{41}_{41}(u)=
4u(u+1),
$$
$$
{\mathfrak{R}}^{12}_{21}(u)={\mathfrak{R}}^{13}_{31}(u)=
{\mathfrak{R}}^{21}_{12}(u)={\mathfrak{R}}^{24}_{42}(u)=
{\mathfrak{R}}^{31}_{13}(u)={\mathfrak{R}}^{34}_{43}(u)=
{\mathfrak{R}}^{42}_{24}(u)={\mathfrak{R}}^{43}_{34}(u)=
2u+3,
$$
$$
{\mathfrak{R}}^{14}_{41}(u)={\mathfrak{R}}^{23}_{32}(u)=
{\mathfrak{R}}^{32}_{23}(u)={\mathfrak{R}}^{41}_{14}(u)=
4u+3,
$$
$$
{\mathfrak{R}}^{14}_{23}(u)={\mathfrak{R}}^{23}_{14}(u)=
{\mathfrak{R}}^{32}_{41}(u)={\mathfrak{R}}^{41}_{32}(u)=
2u,\quad{\mathfrak{R}}^{14}_{32}(u)={\mathfrak{R}}^{23}_
{41}(u)={\mathfrak{R}}^{32}_{14}(u)={\mathfrak{R}}^{41}_
{23}(u)=-2u.
$$
It can be rewritten in the  following form
\be\lb{r44}
{\mathfrak{R}}^{\a_1\a_2}_{\b1\b_2}(u)=2u(2u+3)\d^{\a_1}
_{\b_1}\d^{\a_2}_{\b_2}+(2u+3)\d^{\a_1}_{\b_2}\d^{\a_2}_
{\b_1}-2uK^{\a_1\a_2}_{\b1\b_2},
\ee
where the elements of the matrix $K=4P_{\frac52}$ can be written as
\be\lb{k}
K^{\a_1\a_2}_{\b1\b_2}=\e^{\a_1\a_2}\e_{\b1\b_2},\quad
\quad\e^{\a_1\a_2}=(-1)^{\a_1}\d^{\a_1+\a_2,5},\quad\quad
\e_{\b_1\b_2}=(-1)^{\b_1}\d_{\b_1+\b_2,5}.
\ee
In this matrix after rescaling $2u\to u$ one can recognize the $sp(4)$
vector-vector $R$-matrix (\ref{rzW}) with the parameter $\b=\frac42-(-1)=3$.

We summarize the results on $so(5)$:

\begin{proposition}

The spinorial $R$ matrix with $so(5)$ symmetry coincides with the
ordinary fundamental $R$ matrix with $sp(4)$ symmetry.
The diagonalization of the traces of the corresponding spinorial 
$so(5)$ monodromy and of the ordinary $sp(4)$ monodromy is done 
by the same nested ABA relations. Moreover, the fusion relation  
(\ref{t2t}) allows to treat the diagonalization of the trace of the 
ordinary $so(5)$ monodromy matrix on the basis of the $sp(4)$ nested ABA.
\end{proposition}

The nested ABA for $sp(4)$ has been formulated in \cite{BN}.

\section{Conclusions}

We have presented a new approach to the spinor $R$ matrix
of orthogonal or symplectic symmetry. Comparing with the conventional
approach, the derivation is simpler and the result has the compact form of
the Euler Beta function of the invariant $z$. 

In the orthogonal case, relying on the characteristic polynomial of $z$, we
obtain explicit expressions of the spinorial $R$ matrices of low rank cases.

By the fusion argument we relate the spinor and vector monodromy matrices. 

Studying the low rank examples, we observe coincidences of spinor $R$
matrices with some fundamental $R$ matrices. This implies relations for the 
monodromy matrices, the corresponding RTT algebras and of the Algebraic
Bethe Ansatz  for the spectral problem of traces of spinor and vector
monodromy matrices. 

For the $so(2m)$ cases ($D$ series)  the spinor $R$ matrices  have a
particular simple structure. We have the decomposition into two chiral
parts, where both obey the Yang-Baxter relations. Moreover, the $R$ matrix of
each chiral part is sparse with many zeros and transforms to two
independent blocks.

Coincidence relations between spinor and fundamental $R$ matrices are to be
expected also at higher ranks. The simplicity of the $so(2m)$ case compared
to $so(2m+1)$ will persist.

\vspace{1cm}

\noindent
{\bf Acknowledgments.}

\noindent
The authors are grateful to C. Burdik and O. Navratil for discussions.
They thank the Chech Technical University for hospitality.
The work of D.K. was partially supported by
the Armenian State Committee of Science grant
18T-132 and by Regional Training Network on
Theoretical Physics sponsored by
Volkswagenstiftung Contract nr. 86 260.

\vspace{1cm}

\section{Appendix: Characteristic polynomials}
\setcounter{equation}{0}

Recall the relation  of
the invariant $z=\frac12\c^a_1\c^a_2$ to the conventional spinor
invariants $I_m$ 
\be\lb{im}
I_m=\frac1{m!2^m}\c^{a_1\ldots a_m}_1\c^{a_1\ldots a_m}_2,
\ee
given by the iteration
\be\lb{imzo}
I_{m+1}=zI_m-\frac m4(d-m+1)I_{m-1},\qquad\qquad I_0(z,d)=1,\;\;I_1(z,d)=z.
\ee
It allows to calculate the explicit form the few first invariants:
$$
I_2(z,d)=z^2-\frac d4,
$$

$$
I_3(z,d)=z^3+(\frac12-\frac34d)z,
$$
\be\lb{ii}
I_4(z,d)=z^4+(2-\frac32d)z^2+\frac3{16}d^2-\frac38d,
\ee
$$
I_5(z,d)=z^5+(5-\frac52d)z^3+(\frac32-\frac{25}8d+\frac{15}{16}d^2)z,
$$
$$
I_6(z,d)=z^6+(10-\frac{15}4d)z^4+(\frac{23}2-\frac{105}8d+\frac{45}{16}d^2)z^2
-\frac{15}8d+\frac{45}{32}d^2-\frac{15}{64}d^3,
$$
$$
I_7(z,d)=z^7+(\frac{35}2-\frac{21}2d)z^5+(49-\frac{315}2d+\frac{105}{16}d^2)z^3
+(\frac{45}4-\frac{441}{16}d+\frac{105}8d^2-\frac{105}{64}d^3)z.
$$

The algebra $\mathcal{C}$ implies in the orthogonal case 
the vanishing of $I_{m}(z,d) $ starting from $m=d+1$.
Thus  we obtain the characteristic polynomial in $z$, 
\be \label{Wd}
 W_d(z) = I_{d+1}(z,d) = 0. 
\ee

\subsection{Even-dimensional space}

 Consider rotations in an even-dimensional $d=2m$ Euclidean space. 
$so(d)$ has the $2^m$-dimensional spinor representation $\mathcal{S}$. 
The invariant $z=\frac12\sum_{a=1}^{2m}\c_1^a\c_2^a\equiv\sum_{a=1}^{2m}
z^{(a)}$ acts in $2^{2m}(=2^m\times2^m)$-dimensional space $\mathcal{S}_1
\otimes\mathcal{S}_2$.  It is a sum of 
$2m$ mutually commuting operators $z^{(a)}$ which have $2^{2m}$ common 
eigenvectors $\xi_i$, $i=1,\ldots 4^m$. The eigenvectors $\xi_i$ span 
the $2^{2m}$-dimensional space $\Sigma=\mathcal{S}
\otimes\mathcal{S}$ which decomposes into the direct 
sum of  two $2^m$ dimensional subspaces 
annihilated  by projectors $\Pi_+$ and $\Pi_-$, correspondingly. Here 
$\Pi_\pm=\frac12(1\pm\c_1^{2m+1}\c_2^{2m+1})$, which means that all 
these $2^{2m}$ vectors are the eigenvectors of the operator $z^{(2m+1)}
=\frac12\c_1^{2m+1}\c_2^{2m+1}$ with eigenvalue $\frac12$ or $-\frac12$.
Note that due to the rotational symmetry, each of the operators $z^{(a)}$, 
$a=1,\ldots 2m$ also has the eigenvalue $\frac12$ on half of all 
vectors of $\Sigma$ and the eigenvalue $-\frac12$ on the other half of 
vectors of $\Sigma$.

The eigenvalue of $z$ on the vector $\xi_i$, which is a root 
of the characteristic polynomial, is given by the sum of the eigenvalues 
of $z^{(a)}, a=1, ..., 2m$ on $\xi_i$.  The largest is obtained in the
configuration of "all spins up" and is equal to $m =2m \cdot \half$. This
configuration is encountered once, i.e. here is no degeneracy. 
The largest negative eigenvalue arises from "all spins down", is equal
to $- m= 2m (-\half)$ and is not degenerate either. The next configuration with
"one spin down and all others up" is encountered  $2m=\left(\ba{cc}2m\\1\ea\right)$
times, and the corresponding eigenvalue is $m-1$. Continuing this counting,
we have in particular the eigenvalue $0$ of highest degeneracy 
$\left(\ba{cc}2m\\m\ea\right)$. Thus we obtain the eigenvalues of $z$
equal to $k=-m, -m+1, ..., +m$ and the dimension of the eigenspace
corresponding to $k$ equal to $ \left(\ba{cc}2m\\m+k\ea\right)$.

In this way we obtain the characteristic polynomial of the form
$$
 W_d = z\prod_{k=1}^m(z-k)(z+k)=z\prod_{k=1}^m(z^2-k^2)=0,
$$ 
and we check (by the binomial formular for $(1+1)^{2m}$)
that the number of eigenvectors and eigenvalues 
accounting for their degeneracy equals $2^{2m}$, i.e. the dimension of
$\Sigma= \mathcal{S}_1 \otimes  \mathcal{S}_2 $.

\subsection{The permutation operator}
The operator ${\mathcal{P}}_{12}$ permutes the factors in the  product
$\mathcal{S}_1\otimes \mathcal{S}_2$. It commutes with each $z^{(a)}$, thus 
has common  eigenvectors with them, i.e. the whole representation space 
$\Sigma=\mathcal{S}_1\otimes \mathcal{S}_2$
according to action of ${\mathcal{P}}_{12}$ is divided into the sum
of the symmetric and antisymmetric subspaces $\Sigma=\Sigma_+
\oplus\Sigma_-$.  The eigenspaces $\Sigma_+$ and $\Sigma_-$ of 
 ${\mathcal{P}}_{12}$ have the dimensions $(2^m+1)
2^{m-1}$ and $(2^m-1)2^{m-1}$, respectively.

Like any invariant operator acting on $\Sigma$ the operator ${\mathcal{P}}
_{12}$ for $so(2m)$  has the spectral expansion over the 
projection operators on the eigenspaces of $z$
\be\lb{pperm}
{\mathcal{P}}_{12}=\sum_{k=-m}^m(-1)^{k(k-1)/2}P_k.
\ee
Note that $P_0$ and $P_1$ always contribute with plus sign. 

This formula is closely related to the well known formula
\be\lb{cperm}
{\mathcal{P}}_{12}=\sum_{k=0}^{2m}(-1)^{k(k-1)/2}\frac1{2^m(k!)^2}
\c_1^{a_1\ldots a_k}\c_2^{a_1\ldots a_k}=
\ee
$$
=\frac1{2^m}\Big({\mathbb{I}}_{12}+\frac1{(1!)^2}\c_1^a\c_2^2-\frac1{(
2!)^2}\c_1^{ab}\c_2^{ab}-\frac1{(3!)^2}\c_1^{abc}\c_2^{abc}+\ldots\Big),
$$
as well as with the dimension formula
\be\lb{dsig}
dim(\Sigma_\pm)=rank(\frac12(1\pm{\mathcal{P}}_{12}))=
\sum_{k=0}^{2m}\frac12\Big(1\pm(-1)^{(m+k)(m+k-1)/2}\Big)
\left(\ba{cc}2m\\ k\ea\right)=
\ee
$$
=(2^m\pm1)2^{m-1}.
$$

In order to emphasize the pairwise appearance of the projection
operators in the even-dimensional case we rewrite the expression (\ref{pperm})
for the permutation as
\be\lb{pperm1}
{\mathcal{P}}_{12}=\sum_{k=0}^{m+1}(-1)^k(P_{2k}+P_{2k+1}).
\ee

\subsection{Odd-dimensional space}
If the dimension of the space is odd $d=2m+1$, the dimension of
$\mathcal{S}_1
\otimes \mathcal{S}_2$ and the
number of $z$ eigenvectors remains the same $2^{2m}$, but the operator $z$ 
contains an additional term which shifts its eigenvalues by one half 
and breaks the symmetry between
positive and negative eigenvalues. With the shifted eigenvalues the relation
 $W_d=0$ (\ref{Wd}) holds.
Further, $\e_{a_1, ..., a_{2m+1}}$ causes
relations between even and odd invariants, $ I_k(z) = const \ I_{2m+1-k}(z) $. 
As a consequence we obtain the reduction of $W_d$ to 
\be \label{Wdt}
 \tilde W_d = \prod_{k=0}^m(z-(-1)^k\frac{2k+1}2)=0.
\ee  
The product runs over only $m+1$ of the $2m+1$ roots of $z$.

Consider, how the eigenspaces of $z$ change when increasing the dimension
from $d=2m$ to $d=2m+1$.
The adjacent $z$ multiplets  for $d=2m$ symmetric under the
permutation $\mathcal{P}_{12}$ (e.g. $P_0$ and $P_1$) are unified into a
 multiplet for $d=2m+1$ with the eigenvalue given by arithmetical 
mean ($P_{\frac12}$) by addition of $\frac12$ or $-\frac12$. Similarly, the
adjacent antisymmetric multiplets (say $P_{-2}$ and $P_{-1}$) are  unified
(into $P_{-\frac32}$). Then the dimension formula (\ref{dsig}) holds due to
the Pascal triangle recurrence relation
$$
\left(\ba{cc}2m+1\\ k\ea\right)=\left(\ba{cc}2m\\ k\ea\right)+\left(\ba{cc}2m\\ 
k-1\ea\right),
$$
and accordingly the characteristic polynomial is given by $\tilde W_d$
(\ref{Wdt}).

The dimensions of the subspaces corresponding to the roots of $\tilde W_d$
are given by $\left(\ba{cc}2m+1\\2k\ea\right)$, and their sum is the
dimension of $\Sigma$, 
$$
\sum_{k=0}^m\left(\ba{cc}2m+1\\2k\ea\right)=2^{2m}.
$$
Here we took into account the binomial fomula for $(1+1)^{2m+1}$ and for
$(1-1)^{2m+1}$.

Consider, how the eigenspaces of $z$ change  if we step up in dimension
further from
$d=2m+1$ to $d=2m+2$. The dimension of the spinor space $\mathcal{S}$
is doubled. $\Sigma_{2m+1}$ is separated 
into subspaces even or odd under the permutation ${\mathcal{P}_{12}} $.
The  ${\mathcal{P}_{12}} $ even  subspace of  $\Sigma_{2m+1}$ 
corresponding to the $z$ eigenvalue $\frac12$ 
is translated to the two subspaces of $z$  in $\Sigma_{2m+2}$ 
 with $z$ eigenvalues $0= \frac12- \frac12$ and $1= \frac12+ \frac12 $ 
 Similarly, the  ${\mathcal{P}_{12}} $ odd  subspaces with the
$z$ eigenvalues
$-\frac32$ 
or $z=\frac52$ are each translated to the two subspaces 
of $z$ eigenvalues $-1=  -\frac32+\frac12$, $-2= -\frac32-\frac12 $ or
$2= \frac52- \frac12$, $3=\frac52+ \frac12$.

\end{document}